\documentclass[11pt]{article}
\usepackage{amsmath,amssymb}
\usepackage{graphicx}
\usepackage{array}
\usepackage[ruled,vlined]{algorithm2e}
\usepackage{multirow}
\usepackage{hhline}

\newtheorem{theorem}{Theorem}[section]

\newtheorem{definition}[theorem]{Definition}
\newtheorem{remark}[theorem]{Remark}

\numberwithin{equation}{section}

\begin{document}

\title{Overlapping community detection algorithms using Modularity and the cosine}
\author{Duy Hieu DO and Thi Ha Duong PHAN\\
Institute of Mathematics\\Vietnam Academy of Science and Technology}
\date{\empty}
\maketitle
Email: ddhieu@math.ac.vn (Do Duy Hieu), phanhaduong@math.ac.vn (Phan Thi Ha Duong).
\begin{abstract}
The issue of network community detection has been extensively studied across many fields. Most community detection methods assume that nodes belong to only one community. However, in many cases, nodes can belong to multiple communities simultaneously.This paper presents two overlapping network community detection
algorithms that build on the two-step approach, using the extended modularity and cosine function. The applicability of our algorithms extends to both undirected and directed graph structures. To demonstrate the feasibility and effectiveness of these algorithms, we conducted experiments using real data.
\end{abstract}

\section{Introduction}

In recent years, many studies have focused on network systems such as social networks, biological networks, and technological networks \cite{new1, new2}. One of the focal issues of those studies is about the community detection problem \cite{bs18,2-out,bs29,bs38}.

 Actually the majority of methods assume that nodes belong to only one community, however in many cases, nodes can participate in multiple communities simultaneously, making the problem more challenging. Some authors have made significant efforts to characterize communities with overlapping nodes, as evidenced by recent papers such as \cite{Peel2017, Peixoto2020, Jebabli2018, Harenberg2014}. Nevertheless truly effective algorithms remain a subject requiring further research and demanding new methods.


\subsection{Overlapping community detection algorithms}

We employ the concept of constructing overlapping communities in two steps: the first step involves using an algorithm to partition the graph into disjoint  communities, while the second step entails examining which communities each vertex can potentially belong to.

We are interested in two following algorithms.

First, we are interested in the algorithm introduced in \cite{main}, and temporarily call it \textbf{Parameterized Overlap Algorithm} (or  \textbf{Paramet. Overlap} for short). For this algorithm, the authors propose a simple approach to identify overlapping communities in a graph $G=(V, E)$, where $V=\{v_1, v_2, ..., v_n\}$ represents $n$ nodes and $E \subseteq V \times V$ represents $m$ edges. Let $C = \{C_1; C_2;...; C_k\}$ be a family of subsets of nodes (called also cluster or community) that covers $V$. One then decides a vertex $v$ will be added in a cluster $C_j$ if their belonging coefficient $F_{vc}$ is greater than or equal to a given threshold parameter $\theta$. The  belonging coefficients is defined as follows.

\begin{equation}\label{ptvh0}
     F_{v C_j} =\begin{cases}
     1 \,\,\,\, if \, \frac{\sum_{(v, u) \in E} \chi_{u C_j}}{d_v} \geq \theta,\\
     0\,\,\,\, otherwise.
     \end{cases}
\end{equation}
where $\chi_{u C_j}$ is equal to $1$ if $u \in C_j$ and $0$ otherwise and $d_v$ is the degree of vertex $v$. 

Thus, the belonging coefficient $F_{vC_j}$  is proportional to the number of adjacent edges of $v$ belonging to the cluster $C_j$. As the value of $\theta$ increases, the degree of overlapping between the communities also increases. 




The second algorithm was introduced in \cite{main2}. We temporarily call this \textbf{Module Overlap Algorithm} because the main idea of this algorithm is based on the objective of increasing the modularity $Q_0$  step by step. For instance, the Modularity $Q_0$ for the overlapping community is defined as follows:
$$
Q_0=\frac{1}{2 m} \sum_{C_j \in C} \sum_{u, v \in V} \alpha_{ uC_j} \alpha_{ vC_j}\left(A_{u v}-\frac{d_u d_v}{2 m}\right),
$$
where 
\begin{equation}\label{pt6}
\alpha_{uC_j}=\frac{d_{uC_j}}{\sum_{C_j \in C} d_{uC_j}}, \mbox{ with } d_{uC_j}=\Sigma_{v \in C_j} A_{u v}.
\end{equation}

\subsection{Random walk on graphs}

Let $G = (V, E)$ be a directed graph with $n$ vertices and $m$ edges, define $A$ the
adjacency matrix of $G$.  For $i = 1, 2, . . . , n$, we define $d^{out}_i$ the out-degree of vertex $i$, and $d^{in}_i$ the in-degree of vertex $i$.

It is worth noting that undirected graphs can be seen as a special case of directed graphs,
where the adjacency matrix is symmetric and the out-degree and in-degree of each vertex are equal.

For conciseness, in the following, we will use the out-degree of a vertex as its degree, denoted by $d_i$ unless stated otherwise.

A random walk on $G$ is a process that starts at a given vertex and moves to another vertex
at each time step, the next vertex in the walk is chosen uniformly at random from among the neighbors of the current vertex.
The matrix $P = [p_{ij}]_{i,j=\overline{1,n}}$ represents the transition probabilities of a
Markov chain associated with a random walk on graph $G$. 
At each vertex $i$, the random walk
can move to vertex $j$ with probability  $p_{ij} = a_{ij} / d_i$ if $(i, j) \in E$. Then $P = D^{-1}A$ where $D$ being the  diagonal matrix $D$ using the vertex out-degrees. Moreover $P^t$ represents the transition probabilities of this random walk after $t$ steps.

 We assume that $G$ is strongly
connected, meaning there is a directed path from any vertex $i$ to every other vertex $j$.
 According to the convergence theorem for finite Markov chains, the associated transition matrix $P$ satisfies
$\lim_{k \rightarrow \infty}P =P_{\infty}$, where $(P_{\infty})_{ij} = \phi_j$, the $j$ th component of the unique stationary distribution $\phi=(\phi_1,\phi_2,...,\phi_n)$.

For an undirected graph $G$ where $a_{ij} = a_{ji}$, we have $d_i = d^{out}_i = d^{in}_i$, and $\phi_i = d_i/2m$ for all vertex $i$.

\subsection{Our contribution}
In this paper, each algorithm consists of two steps. In the first step, we use existing community detection algorithms, such as the Hitting times Walktrap algorithm [11], Walktrap algorithm [34], or the Louvain algorithm [37], to find disjointed communities for the network. In the
second step, we determine whether a vertex belongs to a community or not by using modularity or cosine functions. 

\begin{itemize}

\item In Section \ref{modul}, we introduce a concept called Theta-Modularity, an extension of regular Modularity. 
The criteria for a vertex to belong to a community is that the number of edges between them must be sufficiently large and dependent on the total degree of the vertices in that community.

\item  In Section \ref{cosin}, we identify each vertex as a vector and define the center of a cluster to be the vertex corresponding to the vector the average coordinates of the vertices in that cluster. Finally, we will propose an overlapping algorithm based on the idea that vertices of the same cluster will create a small angle with the cluster center; in other words, the cosine of that vertex and the cluster center are more significant than a constant $\theta $.

\end{itemize}
Finally, to assess the effectiveness and rationality of our algorithms, we will conduct experiments to compare and evaluate the clustering results of our algorithms with other algorithms
Specifically, in Subsection \ref{E_vohuong}, we will compare two of our algorithms for undirected graphs, namely the Parameterized Modularity Overlap Algorithm and the Cosine Overlap Algorithm, with two other algorithms Parameterized Overlap Algorithm \cite{main} and the Module Overlap Algorithm \cite{main2}. 
Additionally, we will perform experiments on real datasets and compare the algorithms based on Modularity. In Subsection \ref{E_cohuong}, we will apply our directed graph algorithms to randomly generated graphs and compare the results with the clustering generated by the graph generation model.

\section{Overlapping community detection using modularity}\label{modul}

In this section, we will use the modularity function to determine whether a vertex belongs to a community or not. The classic modularity function evaluates the difference between the real number of edges between two vertices and the expected number of edges between them. Our new approach is to introduce a threshold for the expected number of edges in order  to make the evaluation more flexible. This new parameterized modularity function will be applied to undirected graphs in section 2.1, and to directed graphs in section 2.2 where the expected number is calculated based on the degree of the vertices. 
A breakthrough in section 2.3 is that the parameterized modularity function for directed graphs will be defined
based on the stationary distribution of a random walk on the graph. This provides a more advanced approach to determining community membership.

\subsection{Overlapping community detection for undirected graphs}\label{2.1}

The expected number of edges falling between two vertices $u$ and $v$ in the
configuration model is equal to $d_ud_v/2m$,
then the actual-minus-expected edge count for the vertex pair $(u,v)$ is $A_{uv} -d_ud_v/2m$. Suppose $C=\{C_1, C_2,..., C_k\}$ is a cover of the vertices of the undirected graph $G$, the modularity $Q$ (as defined in \cite{10}) is then equal to 
\begin{equation}\label{ptQ0}
Q=\frac{1}{2 m} \sum_{u ,v\in V}\left(A_{u v}- \frac{d_u d_v}{2 m}\right) \delta\left(C_u, C_v\right)\\
= \frac{1}{2 m} \sum_{C_j\in C} \sum_{u ,v\in C_j}\left(A_{u v}- \frac{d_u d_v}{2 m}\right).
\end{equation}
where $\delta\left(C_u, C_v\right)$ is $1$ if $u$ and $v$ are in the same community, and $0$ otherwise. 

This modularity illustrates the criteria that $u$ and $v$ belong to the same cluster if the real number of edges between them is greater than the expected number of edges between them. However, we find many practical problems when dividing clusters by data, depending on the goal the required criteria is more flexible. Therefore, we propose a new modularity with theta coefficient as follows.
\begin{equation}\label{ptQ}
Q(\theta)=\frac{1}{2 m} \sum_{C_j\in C} \sum_{u ,v\in C_j}\left(A_{u v}-\theta \frac{d_u d_v}{2 m}\right).
\end{equation}
Thís new modularity means that two vertices $u$ and $v$ belong to the same cluster if the number of edges between $u$ and $v$ is more significant than $\theta$-times the expected number of edges between them.

 We observe that the modularity of clustering of graph $G$ can be expressed as the sum of the modularity of each cluster, as shown in formula \ref{ptQ}. This implies that if we add a vertex to a cluster, only the modularity value of that specific cluster will be affected. As a result, for every community $C_j$ and vertex $u \in V(G) \setminus  C_j$, the modularity of cluster $C_j$ will change by an amount when we add vertex $u$ to it, and the following formula can calculate it.  
\begin{equation}
\Delta Q(\theta)_{u,C_j} = \dfrac{1}{2m}\sum_{w\in C_j  }\left( A_{uw}-\theta\dfrac{d_ud_w}{2m}\right).
\end{equation}
From the above comment, we propose that vertex  $u$ will be added to the community $C_j$ if $\Delta Q(\theta)_{u, C_j}$ is positive:
\begin{equation}\label{pt1.22}
 \dfrac{1}{2m}\sum_{w\in C_j  }\left( A_{uw}-\theta\dfrac{d_ud_w}{2m}\right) >0,
\end{equation}
which corresponds to
\begin{equation}\label{ptvh1}
\sum_{w\in C_j  } \frac{A_{uw}}{d_u}> \theta\sum_{w\in C_j} \dfrac{d_w}{d}.
\end{equation}
\begin{remark}
From (\ref{ptvh0}) and (\ref{ptvh1}), we have remarked that vertex  $u$ belongs to the community $C_j$ if the number of edges between $u$ and $C_j$ is large enough. However, in the equation (\ref{ptvh0}), the number of edges between $u$ and $C_j$ must always be greater than a fixed $\theta$ constant. It is independent of the properties of the community $C_j$. Our method is more reasonable because it depends not only on the $\theta$ coefficient but also on the characteristics of the community $C_j$. Specifically, it depends on the sum of the degrees of the vertices in the community $C_j$.
\end{remark}
From there, we propose the overlap detection algorithm described in \textbf{Algorithm} \ref{alg:modul-Q-overlap0}. We will call this the \textbf{Parameterized Modularity Overlap Algorithm} for undirected graphs (or  \textbf{Paramet. Modul.} for short). 
\\

\begin{algorithm}[H]
\SetAlgoLined
\KwIn{An undirected graph $G$ and a threshold value $\theta$}
\KwOut{Clusters of vertices with overlapping communities}
Apply the Louvain algorithm \cite{Louvain} to obtain the initial clustering of $G$ into communities $C_1, C_2,..., C_k$\;
\Repeat{No communities $C_j$ meet the condition}{
    \ForEach{vertex  $u$}{
        \ForEach{community $C_j$ adjacent to $u$ and not containing $u$}{
            \If{$\sum_{w\in C_j  } \frac{A_{uw}}{d_u}> \theta\sum_{w\in C_j} \dfrac{d_w}{d}$}{
                Add $u$ to $C_j$\;
            }
        }
    }
}
\caption{Parameterized Modularity Overlap Algorithm}
\label{alg:modul-Q-overlap0}
\end{algorithm}

In the Algorithm \ref{alg:modul-Q-overlap0}, we have to traverse all the vertices, and for each vertex, we consider all the adjacent communities with it. So the computational complexity of this Algorithm will be $O(k n)$, where $n$ is the number of vertices of the graph, and $k$ is the number of communities.

\subsection{Overlapping community detection for directed graphs using modularity}\label{2.2}

In \cite{p1,p8}, the authors had given the  in/out-degree sequence of directed graph,
in which the probability to have an edge from vertex $v$ to vertex  $u$ is determined by $d^{in}_u d^{out}_v$, where $d^{in}_u$ and $d^{out}_v$ are the in- and out-degrees of the vertices. Suppose $C=\{C_1, C_2,..., C_k\}$ is a cover of $G$, the modularity $Q$ is defined as.
\begin{equation}\label{Qcohuong}
    Q^d=\frac{1}{m} \sum_{C_j\in C} \sum_{u,v\in C_j}\left( A_{uv} - \dfrac{d^{in}_ud^{out}_v}{m} \right),
\end{equation}
where $A_{uv}$ is defined conventionally to be $1$ if there is an edge from $v$ to $u$ and zero otherwise. Note
that indeed edges $v \rightarrow u$ make larger contributions to this
expression if $d^{in}_u$ and/or $d^{out}_v$
is small.

Similar to the case of undirected graphs, we also define Theta-Modularity as follows:
\begin{equation}
    Q^d(\theta)=\frac{1}{m}\sum_{C_j\in C} \sum_{u,v\in C_j}\left( A_{uv} - \theta\dfrac{d^{in}_ud^{out}_v}{m} \right).
\end{equation}
 Then for each community $C_j$, we will consider the vertices $u\in V(G)\setminus  C_j$. We notice that if a vertex  $u$ added to the community $C_j $, the modulus of the cluster $C_j $ changes by an amount of
\begin{equation}
\Delta Q^d(\theta)_{u,C_j} =  \frac{1}{m}\sum_{w\in C_j}\left( A_{u w} - \theta\dfrac{d^{in}_ud^{out}_w}{m} \right)+\frac{1}{m}\sum_{w\in C_j}\left( A_{ w u} - \theta\dfrac{d^{in}_w d^{out}_u}{m} \right).
\end{equation}
From the above comment, we propose that vertex  $u$ belongs to the community $C_j$ if $\Delta Q^d(\theta)_{u, C_j}$ is positive:
\begin{equation}\label{pt1.2}
\frac{1}{m}\sum_{w\in C_j}\left( A_{u w} - \theta\dfrac{d^{in}_ud^{out}_w}{m} \right)+\frac{1}{m}\sum_{w\in C_j}\left( A_{ w u} - \theta\dfrac{d^{in}_w d^{out}_u}{m} \right) >0.
\end{equation}
equivalent to
\begin{equation}\label{ptch1}
\sum_{w\in C_j}\left( A_{u w}+A_{ w u}\right)  >\theta\sum_{w\in C_j}\left(  \dfrac{d^{in}_ud^{out}_w}{m} + \dfrac{d^{in}_w d^{out}_u}{m} \right),
\end{equation}
\begin{remark}
From (\ref{ptch1}), we have remarked that vertex  $u$ belongs to the community $C_j$ if the total number of edges from $u$ to $C_j$ and the number of edges from $C_j$ to $u$ is large enough. 
\end{remark}
From there, we propose the overlap detection algorithm described in \textbf{Algorithm} \ref{alg:modul-Q-overlap-directed}. We will call this the \textbf{Directed Parameterized d-Modularity Overlap Algorithm} for directed graphs (or  \textbf{Di-Paramet. d-Modul.} for short)
\\

\begin{algorithm}[H]
\SetAlgoLined
\KwIn{A directed graph $G$ and a threshold value $\theta$}
\KwOut{Clusters of vertices with overlapping communities}
Apply Louvain algorithm \cite{louvain_cohuong} to obtain the initial clustering of $G$ into communities $C_1, C_2,..., C_k$\;
\Repeat{No communities $C_j$ meet the condition}{
    \ForEach{vertex  $u$}{
        \ForEach{community $C_j$ adjacent to $u$ and not containing $u$}{
            \If{$\sum_{w\in C_j}\left( A_{u w}+A_{ w u}\right)  >\theta\sum_{w\in C_j}\left(  \dfrac{d^{in}_ud^{out}_w}{m} + \dfrac{d^{in}_w d^{out}_u}{m} \right)$}{
                Add $u$ to $C_j$\;
            }
        }
    }
}
\caption{Directed Parameterized d-Modularity Overlap Algorithm}
\label{alg:modul-Q-overlap-directed}
\end{algorithm}
Similar to the Algorithm \ref{alg:modul-Q-overlap0}, the Algorithm \ref{alg:modul-Q-overlap-directed}  also has a computational complexity of $O(k n)$, where $n$ is the number of vertices of the graph, and $k$ is the number of communities.

\subsection{Overlapping community detection for directed graphs using the stationary
distribution}\label{2.3}

Many modularities have been proposed for directed graphs, such as the Modularity in Formula \ref{Qcohuong}. In many cases, those modularity proposals will lose the essential properties of directed graphs. Therefore, in \cite{cohuong}, we proposed a definition of modularity for directed graphs based on random walks and stationary distribution, which is a natural extension of modularity on undirected graphs. In detail, for a cover $C=\{C_1, C_2,..., C_k\}$ of the directed graph $G$, our proposed modularity is the following.

\begin{equation}\label{ourQ}
Q^{sd}=\sum_{uv}\left(P_{vu}\phi_{v}-\phi_{u}\phi_{v}\right)\delta_{C_{u}C_{v}},
\end{equation}
where $P_{vu}$ is the transition probability of the random walk process from
$v$-th vertex to $u$-th vertex and  $\phi=\left(\phi_{1},\phi_{2},...,\phi_{n}\right)$ is the stationary distribution stationary.

We also have the $\theta$ modularity version  as follows.
\begin{equation}
Q^{sd}(\theta)=\sum_{uv}\left(P_{vu}\phi_{v}-\theta\phi_{u}\phi_{v}\right)\delta_{C_{u}C_{v}}.
\end{equation}
and then
\begin{equation}
Q^{sd}(\theta)=\sum_{C_j\in C} \sum_{u,v\in C_j}\left(P_{vu}\phi_{v}-\theta\phi_{u}\phi_{v}\right).
\end{equation}
 Then for each community $C_j$, we will consider the vertices $u\in V(G)\setminus  C_j$: if $u$ is added to the community $C_j $ the modularity of the cluster $C_j $ changes by an amount of
\begin{equation}\label{deltaQ}
\Delta Q^{sd}(\theta)_{u, C_j}=\sum_{w\in C_j  }\left(\phi_u P_{uw} -\theta \phi_u\phi_w  \right) + \sum_{w\in C_j }\left( \phi_w P_{wu}- \theta\phi_u\phi_w  \right).
\end{equation}
We also propose that vertex  $u$ belongs to community $C_j$ if $\Delta Q^{sd}(\theta)_{u, C_j}$ is positive, that means:
\begin{equation}\label{pt2.2}
\sum_{w\in C_j}\left(\phi_u P_{uw} -\theta \phi_u\phi_w  \right) + \sum_{w\in C_j}\left( \phi_w P_{wu}- \theta\phi_u\phi_w  \right)  >0,
\end{equation}
which is equivalent to
\begin{equation}\label{ptch2}
\sum_{w\in C_j}\left(\phi_u  P_{uw} + \phi_w P_{wu}  \right) > 2\theta\sum_{w\in C_j}\phi_u\phi_w.
\end{equation}
\begin{remark}
The formula (\ref{ptch2}) means that vertex  $u$ belongs to the community $C_j$ if the sum of the probabilities from vertex  $u$ to the community $C_j$ and the probabilities from community $C_j$ to vertex  $u$ is large enough.
\end{remark}
From there, we also propose the \textbf{Algorithm} \ref{alg:modul-Q-overlap-directed2} that we call \textbf{Directed Parameterized sd-Modularity Overlap Algorithm} for directed graphs (or  \textbf{Di-Paramet. sd-Modul.} for short).
\\

\begin{algorithm}[H]
\SetAlgoLined
\KwIn{A directed graph $G$ and a threshold value $\theta$}
\KwOut{Clusters of vertices with overlapping communities}
Apply Louvain algorithm \cite{louvain_cohuong} to obtain the initial clustering of $G$ into communities $C_1, C_2,..., C_k$\;
\Repeat{No communities $C_j$ meet the condition}{
    \ForEach{vertex  $u$}{
        \ForEach{community $C_j$ adjacent to $u$ and not containing $u$}{
            \If{$
\sum_{w\in C_j}\left(\phi_u  P_{uw} + \phi_w P_{wu}  \right) > 2\theta\sum_{w\in C_j}\phi_u\phi_w$}{
                Add $u$ to $C_j$\;
            }
        }
    }
}
\caption{Directed Parameterized sd-Modularity Overlap Algorithm}
\label{alg:modul-Q-overlap-directed2}
\end{algorithm}

Algorithm \ref{alg:modul-Q-overlap-directed2} requires computing the stationary distribution and using two loops similar to Algorithm 1 and Algorithm 2. Various efficient computation algorithms exist to calculate the stationary distribution, such as the one presented in \cite{hittinganddis}, which has a computational complexity of $\tilde{O} (m^{3/4}n+m n^{2/3})$, where the $\tilde{O}(n)$ notation suppresses polylogarithmic factors in $n$. Therefore, the total computational complexity of Algorithm 3 is the sum of $O(kn)$ and $\tilde{O} (m^{3/4}n+m n^{2/3})$, which equals $\tilde{O} (m^{3/4}n+m n^{2/3})$.

\begin{remark}
The equation (\ref{pt1.22}) for undirected graphs can be rewritten as follows.
\begin{equation}\label{pt1.222}
 \dfrac{1}{2m}\sum_{w\in C_j  }\left( A_{uw}-\theta\dfrac{d_ud_w}{2m}\right) + \dfrac{1}{2m}\sum_{w\in C_j  }\left( A_{wu}-\theta\dfrac{d_ud_w}{2m}\right) >0.
\end{equation}
and because $A_{uw} =d_u P_{uw}$, $d=2m$ $\phi_u=\frac{d_u}{d}$, then from (\ref{pt1.222}), we get
\begin{equation}\label{pt1.4}
\sum_{w\in C_j}\left(\dfrac{d_u}{d} P_{uw}-\theta\dfrac{d_u}{d} \dfrac{d_w}{d} \right)  + \sum_{w\in C_j}\left(\dfrac{d_w}{d} P_{wu}-\theta\dfrac{d_u}{d} \dfrac{d_w}{d} \right) >0,
\end{equation}
which implies that
\begin{equation}\label{pt11.6}
\sum_{w\in C_j}\left(\phi_u P_{uw}+\phi_w P_{wu}\right)>2\theta \sum_{w\in C_j}\phi_u\phi_w.
\end{equation}

From (\ref{ptch2}) and (\ref{pt11.6}), we observe that the condition for a vertex $u$ to belong to a community $C_j$ that we propose in a directed graphs is a natural extension of that in undirected graphs through the random walk and stationary distribution.

\end{remark}

\section{Overlapping community detection using the cosine}\label{cosin}

In some studies (as seen in \cite{L_kd}), the authors have represented vertices of a graph as vectors in space and defined two vertices belong to the same community when the angle formed by their respective vectors is small. Consequently, the cosine of the angle between them is approximately 1.

Expanding on this concept, we propose the following algorithms: a vertex $u$ belongs to a community $C$ if the cosine of the angle between the vector of $u$ and the vector of the center of $C$ is approximately 1. In this scenario, the vector of the center of $C$ is calculated by averaging the coordinates of all vertices within the community.


\subsection{Overlapping community detection for undirected graphs}\label{3.1}

In \cite{Walktrap}, the authors
noticed that: two vertices $u$ and $v$, that are closed each other,  tend to "see" all the other vertices in the same way, that means 
\begin{equation}\label{ptP}
 P_{uw}^t  \simeq  P_{vw}^t, \mbox{ for all } w.
\end{equation}
Then they defined the distance between them:
$
 R_{uv}(t) :=  \|D^{-1/2}P^t_{u\bullet} -D^{-1/2}P^t_{v\bullet}\|.
$
Inspiring from this idea, we correspond each vertex  $u$ to the vector $D^{-1/2}P^t_{u\bullet}$, 
\begin{equation}\label{pttoado1}
   Coord(u):=D^{-1/2}P^t_{u\bullet}= \left\{d_1^{-1/2} P^t_{u1},d_2^{-1/2}P^t_{u2},...,d_n^{-1/2} P^t_{un}\right\}.
\end{equation}
From equation \ref{ptP}, we can also observe that if two vertices $u$ and $v$ belong to the same community, the angle formed by the two vectors $Coord(u)$ and $Coord(v)$ will be pretty small. In other words
\begin{equation}\label{cosine}
cosin(Coord(u), Coord(v))\simeq 1.
\end{equation}
Because the lengths of vectors are comparable, and using the cosine function provides an explicit evaluation by comparing to 1, we will use Equation \ref{cosine} to determine whether two vertices are in the same cluster or not.
From this comment, for undirected graphs we propose the following \textbf{Algorithm} \ref{alg:cosine_overlap} that we call \textbf{Cosine Overlap Algorithm}.
\\

\begin{algorithm}[H]
\SetAlgoLined
\caption{Cosine Overlap Algorithm}
\label{alg:cosine_overlap}
\KwIn{Undirected graph $G$, parameter $\theta$}
\KwOut{Clusters $C_1, C_2, ..., C_k$}
Apply Louvain algorithm \cite{Louvain} to obtain initial clusters $C_1, C_2, ..., C_k$ of $G$;
\ForEach{cluster $C_j$}{
Calculate the cluster center $Center_j$ as follows:
$
Center_j = \frac{1}{|C_j|}\sum_{u \in C_j} \text{Coord}(u),
$
where $\text{Coord}(u) = D^{-1/2}P^t_{u\bullet}$ is the coordinate vector of node $u$ defined as
\begin{equation}
\text{Coord}(u) = \left\{d_1^{-1/2} P^t_{u1},d_2^{-1/2}P^t_{u2},...,d_n^{-1/2} P^t_{un}\right\}.
\end{equation}
}
\ForEach{cluster $C_j$}{
\ForEach{vertex  $u$}{
\If{ $\text{cosin}(\text{Coord}(u), Center_j) > \theta$
}{
Add vertex  $u$ to cluster $C_j$;
}{
}
}
}
\end{algorithm}

The parameter $\theta$ is dependent on the network's structure and the desired level of overlap. For networks with a well-defined structure, choosing $\theta$ between $0.7$ and $0.8$ is suitable. However, in cases where the network lacks a clear community structure, opting for a value of $\theta$ lower than $0.7$ is more appropriate.

The computational complexity of the Louvain algorithm is $O(n \log n)$. Determining the coordinates of the vertices carries a complexity of $O(n^3)$, while computing the center has a complexity of $O(n)$. The step involving cosine calculations operates at a complexity of $O(kn^2)$. Consequently, the overall computational complexity of Algorithm 4 amounts to $O(n^3)$.

\subsection{Overlapping community detection for directed graphs}
For a strongly connected digraph $G$, let $\Phi^{1/2} = diag[\sqrt{\phi_u}]$.  Yanhua and Z. L. Zhang \cite{dilaplace} defined the normalized digraph Laplacian matrix (Diplacian for short) $\Gamma = [\Gamma_{uv} ]$ for the graph $G$ as follows.
\begin{definition}(\cite[Definition 3.2]{dilaplace})
The Diplacian $\Gamma$ is defined as
\begin{equation}
    \Gamma = \Phi^{1/2} (I-P) \Phi^{-1/2}.
\end{equation}
\end{definition}
In \cite{cohuong}, we perform singular value decomposition on the normalized Laplace matrix $\Gamma = U \Sigma V^T$, where $V=[V_1, V_2,..., V_n]$; and we take $V_k = [V_1, V_2, \dots, V_k]$ such that the coordination of each vertex $u$ is defined as follows: \begin{equation}\label{co22}  Coord(u) = \left(\phi^{-1/2}_uV_1(u), \dots, \phi^{-1/2}_uV_k(u),\phi^{-1/2}_uU_1(u), \dots, \phi^{-1/2}_uU_k(u)\right). \end{equation}
 Similarly to the case of undirected graphs, we can also observe two vertices $u$ and $v$ belong to the same community if the angle formed by the two vectors $Coord(u)$ and $Coord(v)$ is small, which is equivalent to 
$$cosin(Coord(u), Coord(v))\simeq 1.$$ 
From this comment,  for directed graphs, we also propose the \textbf{Algorithm} that we call \textbf{Directed Cosine Overlap Algorithm} (or  \textbf{Di-Cosine
Overlap Algorithm} for short).
\\

\begin{algorithm}[H]
\SetAlgoLined
\caption{Directed Cosine Overlap Algorithm}
\label{alg:cosine_overlap2}
\KwIn{Directed graph $G$, parameter $\theta$}
\KwOut{Clusters $C_1, C_2, ..., C_k$}
Apply our NL‐PCA algorithm \cite{cohuong} to obtain initial clusters $C_1, C_2, ..., C_k$ of $G$;

\ForEach{cluster $C_j$}{
Calculate the cluster center $Center_j$ as follows:
\begin{equation}
Center_j = \frac{1}{|C_j|}\sum_{u \in C_j} \text{Coord}(u),
\end{equation}
where the coordinate vector of node $u$ is defined as
\begin{equation} \label{co2}
Coord(u) = \left(\phi^{-1/2}_uV_1(u), \dots, \phi^{-1/2}_uV_k(u),\phi^{-1/2}_uU_1(u), \dots, \phi^{-1/2}_uU_k(u)\right). \end{equation} 
}
\ForEach{cluster $C_j$}{
\ForEach{vertex  $u$}{
\If{
$
\text{cosin}(\text{Coord}(u), Center_j) > \theta
$
}{
Add vertex  $u$ to cluster $C_j$;
}{
}
}
}
\end{algorithm}
 We will also choose the parameter $\theta$   as in  Algorithm 4 for undirected graphs.  The computational complexity of the NL-PCA algorithm is $O(n^3)$. The coordinates of vertices are already calculated in NL-PCA algorithm. The complexity for computing the center is $O(kn)$. So, the total computational complexity of Algorithm 5 is $O(n^3)$.

\section{Examples}

\begin{figure}
\begin{centering}
\includegraphics[width=0.3\columnwidth]{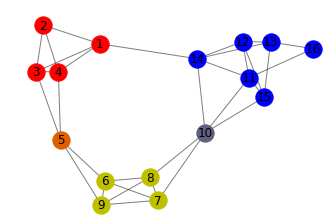}
\par\end{centering}
\caption{Applying the Parameterized Modularity Overlap Algorithm for undirected graphs with $\theta=1$, we obtained three communities are  $C_1=\{1,2,3,4,5\}$,  $C_2=\{5,6,7,8,9,10\}$, and  $C_3=\{10,11,12,13,14,15,16\}$. Vertex $5$ belongs to communities $C_1$ and $C_2$. Vertex $10$ belongs to both communities $C_2$ and $C_3$.}\label{Fig1}
\end{figure}

\begin{figure}
\begin{centering}
\includegraphics[width=0.3\columnwidth]{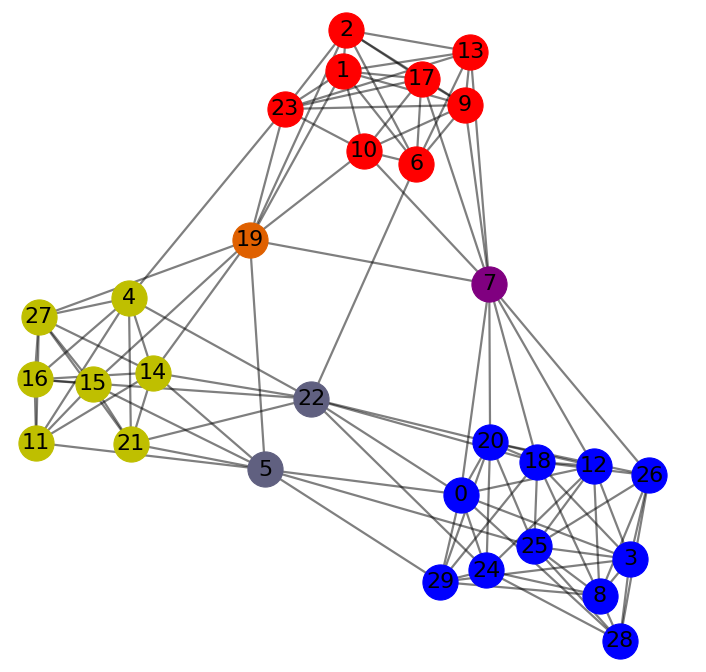}
\par\end{centering}
\caption{Applying the Cosine Overlap Algorithm with $\theta=0.6$, we obtained three communities  $C1=\{4, 5, 11, 14, 15, 16, 21, 22, 27, 19\}$,  $C2=\{1, 2, 6, 7, 9, 10, 13, 17, 19, 23\}$ and  $C3=\{0, 3, 8, 12, 18, 20, 24, 25, 26, 28, 29, 5, 7, 22\}$. Vertex $19$ belongs to communities $C_1$ and $C_2$. Vertex $7$ belongs to both communities $C_2$ and $C_3$, and vertices $5$, $22$ belongs to both communities $C_1$ and $C_3$.}\label{Fig2}
\end{figure}

\begin{figure}
\begin{centering}
\includegraphics[width=0.3\columnwidth]{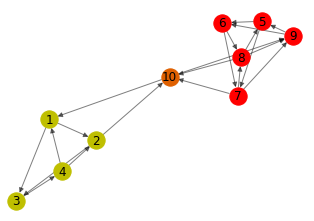}
\par\end{centering}
\caption{Applying the Parameterized Modularity Overlap Algorithm for directed graphs with $\theta=1$, we obtained two communities   $C_1=\{1,2,3,4,10\}$ and $C_2=\{5,6,7,8,9,10\}$. Vertex $10$ belongs to communities $C_1$ and $C_2$.}\label{Fig3_Exa}
\end{figure}

We inlustrate our algorithms for explicit graphs as follows: the Parameterized Modularity Overlap Algorithm  for undirected graph in Figure \ref{Fig1}, the Cosine Overlap Algorithm  for undirected graph in Figure \ref{Fig2}, the Parameterized Modularity Overlap Algorithm for directed graphs in Figure \ref{Fig3_Exa}. 
 
\section{Experiments}

We will evaluate the effectiveness and rationality of our algorithms by conducting experiments to compare and consider the clustering results of our algorithms with other algorithms and the clustering generated by the graph generation model. Specifically: In Subsection \ref{E_vohuong}, we will compare two of our algorithms for undirected graphs, namely the Parameterized Modularity Overlap Algorithm (proposed in Subsection \ref{2.1}) and the Cosine Overlap Algorithm (proposed in Subsection \ref{3.1}), with four other algorithms. In which the two algorithms we introduced in the previous section are the Parameterized Overlap Algorithm \cite{main} and the Module Overlap Algorithm \cite{main2}. Moreover, we also compare our algorithms with two famous algorithms, the Bigclam algorithm \cite{Bigclam} and the Copra algorithm \cite{Copra}.

We will compare these algorithms based on Modularity and ONMI. Additionally, we will perform experiments on real datasets and compare the algorithms based on Modularity. Subsection \ref{E_cohuong}, we will apply our directed graph algorithms to randomly generated graphs and compare the results with the clustering generated by the graph generation model based on ONMI.

\subsection{Evaluating metrics}
 
\subsubsection*{Modularity for undirected graphs with overlapping communities:} Chen et al. \cite{main2} provide the generalized modularity-based belonging function for calculating modularity in undirected graphs. The following equation represents this function:
\begin{equation}\label{Q_over}
Q=\frac{1}{2 m} \sum_{C_j \in C} \sum_{u, v\in C_j}\left(A_{u v}-\frac{d_u d_v}{2 m}\right) f\left(\alpha_{u C_j}, \alpha_{v C_j}\right).
\end{equation}
 where $\alpha_{u C_j}$ is defined as in the Equation (\ref{pt6}). The belonging coefficient function $f\left(\alpha_{u C_j}, \alpha_{v C_j}\right)$ can be the product or average of $\alpha_{u C_j}, \alpha_{v C_j}$. If it is average, it becomes the following equation.
\begin{equation}\label{Quse}
Q=\frac{1}{2 m} \sum_{C_j \in C} \sum_{u ,v\in V}\left(A_{u v}-\frac{d_u d_v}{2 m}\right)\left(\alpha_{u C_j}+\alpha_{u C_j}\right) / 2.
\end{equation}
In this part of the experiment, we will use this modularity to evaluate the clustering quality.

\subsubsection*{The Overlapping Normalized Mutual Information (ONMI):} 
The Overlapping Normalized Mutual Information \cite{ONMI} is a measure used to evaluate the similarity between two clusters or data organizations. It considers both the overlap between clusters and the similarity between labels within the clusters. 
The formula to compute ONMI is as follows: 
\begin{equation}\label{ONMI}
ONMI(A, B) = \dfrac{2\,  MI(A, B)}{H(A) + H(B)},
\end{equation}
where, $MI(A, B)$ represents the Mutual Information (MI) between the two clusters $A$ and $B$.
$H(A)$ and $H(B)$ denote the entropy of clusters $A$ and $B$, respectively.
Mutual Information (MI) measures the dependence between two random variables. It is calculated by summing the joint probabilities of pairs of values for the two variables multiplied by the logarithm of the ratio between the joint probability and the product of the individual probabilities.
Entropy quantifies the uncertainty in a random variable. The entropy of a cluster is calculated by summing the probabilities of the labels within the cluster multiplied by the logarithm of that probability.

\subsection{The random graph model and experiments on random graphs}
\subsubsection{The random graph model}
Evaluating a community detection algorithm is difficult because one needs
some test graphs whose community structure is already known. A classical approach is
to use randomly generated graphs with labeled 
communities. Here we will use this approach and
generate the graphs as follows. \\
\textbf{LFR benchmark graphs:} This random graph generator model creates community-structured graphs with overlapping vertices. Andrea Lancichinetti and Santo Fortunato proposed it in \cite{lfr}. This model generates graphs with many of the same properties as real networks. To create the graphs, we need  the following parameters:
\begin{itemize}
    \item N:		number of nodes.
  \item k:		average degree.
  \item maxk:		maximum degree.
  \item $\mu$:		mixing parameter.
  \item $t_1$:		minus exponent for the degree sequence.
  \item $t_2$:		minus exponent for the community size distribution.
  \item minc:		minimum for the community sizes.
  \item maxc:		maximum for the community sizes.
  \item on:		number of overlapping nodes.
  \item om:		number of memberships of the overlapping nodes.
\end{itemize}
When applying, we take the parameters $t_1=2$, $t_2=1$. This random graph generation model can generate both undirected and directed graphs.

\subsubsection{Experiments on random graphs for undirected graphs}\label{E_vohuong}

In this part of the experiment, each table results from experiments on ten randomly generated graphs using the LFR benchmark graphs mode. We will conduct experiments on all six algorithms for the ten graphs generated. 
For each Algorithm except the Bigclam algorithm, we will perform 20 experiments corresponding to 20 different parameters and select the clustering result with the highest Modularity and ONMI among those experiments. Specifically, the parameters for each Algorithm will be as follows:
\begin{itemize}
    \item 
Parameterized Modularity Overlap Algorithm: we will use the coefficient $\theta=1+0.1t$ with $t\in\{1,2,\dots,20\}$.
 \item Cosine Overlap Algorithm: we will use the coefficient $\theta=0.2+0.035t$ with $t\in\{1,2,\dots,20\}$.
 \item Parameterized Overlap Algorithm: we will use the coefficient $\theta=0.2+0.015t$ with $t\in\{1,2,\dots,20\}$.
 \item Module Overlap Algorithm: we will set the coefficient $B^{U}$ to 0.5 and use the coefficient $B^{L}=0.2+0.015t$ with $t\in\{1,2,\dots,20\}$.
 \item Copra Algorithm: We will apply the Copra algorithm 20 times to each graph and get the best result out of those 20 experiments. To increase the algorithm's accuracy, we choose the parameter $v$ in the algorithm to be the number generated by creating the graph and the parameter $T$ to be 15.
  \item Bigclam Algorithm: To improve the algorithm's accuracy, we choose the parameter $K$ (number of communities), which is the number of communities generated by random graph generation.
\end{itemize}

We will evaluate the efficiency of algorithms using two metrics: modularity (the Modularity we use is the Formula \ref{Quse}) and ONMI.
In each experiment, we will first compare the maximum modularity values obtained from the clustering results obtained by each algorithm. A higher modularity value indicates a more efficient algorithm for forming well-defined communities.
Next, we will compare the clustering results obtained by each algorithm with the original clustering generated by the graph generation (original clustering) based on ONMI. ONMI measures the similarity between two sets of clusters, with a value close to 1 indicating a strong resemblance between the obtained and original clustering. When the obtained clustering matches the original clustering, the ONMI value will be 1.
By employing both modularity and ONMI, we can assess the algorithm's effectiveness in forming cohesive communities and its ability to generate clustering results that closely align with the ground-truth clustering.

In the LFR benchmark graphs model, the coefficient $\mu \in [0,1]$ determines the clarity of the community structure. When $\mu$ is close to 0, the graph exhibits a clear community structure with well-defined communities. On the other hand, when $\mu$ approaches 1, the generated graph shows little to no community structure, with a high level of overlap between communities.
To thoroughly explore the algorithm's behavior, we will test it in each experiment on various values of $\mu$. Specifically, we will experiment with the following values of $\mu$: $\mu \in \{0.05, 0.075, 0.1, 0.2, 0.3, 0.4\}$. By testing the algorithm on these different values of $\mu$, we can analyze its performance under various community structure scenarios, ranging from clear to overlapping communities.

\subsubsection*{Experiment 1 for undirected graphs:} 
We will experiment on ten randomly generated graphs using the LFR benchmark with all the parameters taken with a uniform distribution in the following corresponding intervals: $N \in [400;500]$, $k \in [6;8]$, $maxk\in [8;16]$, $on \in [60;80]$, $om \in [2;5]$, $minc\in [30;50]$, and $maxc \in [50;80]$.

\begin{figure}
\begin{centering}
\includegraphics[width=1\columnwidth]{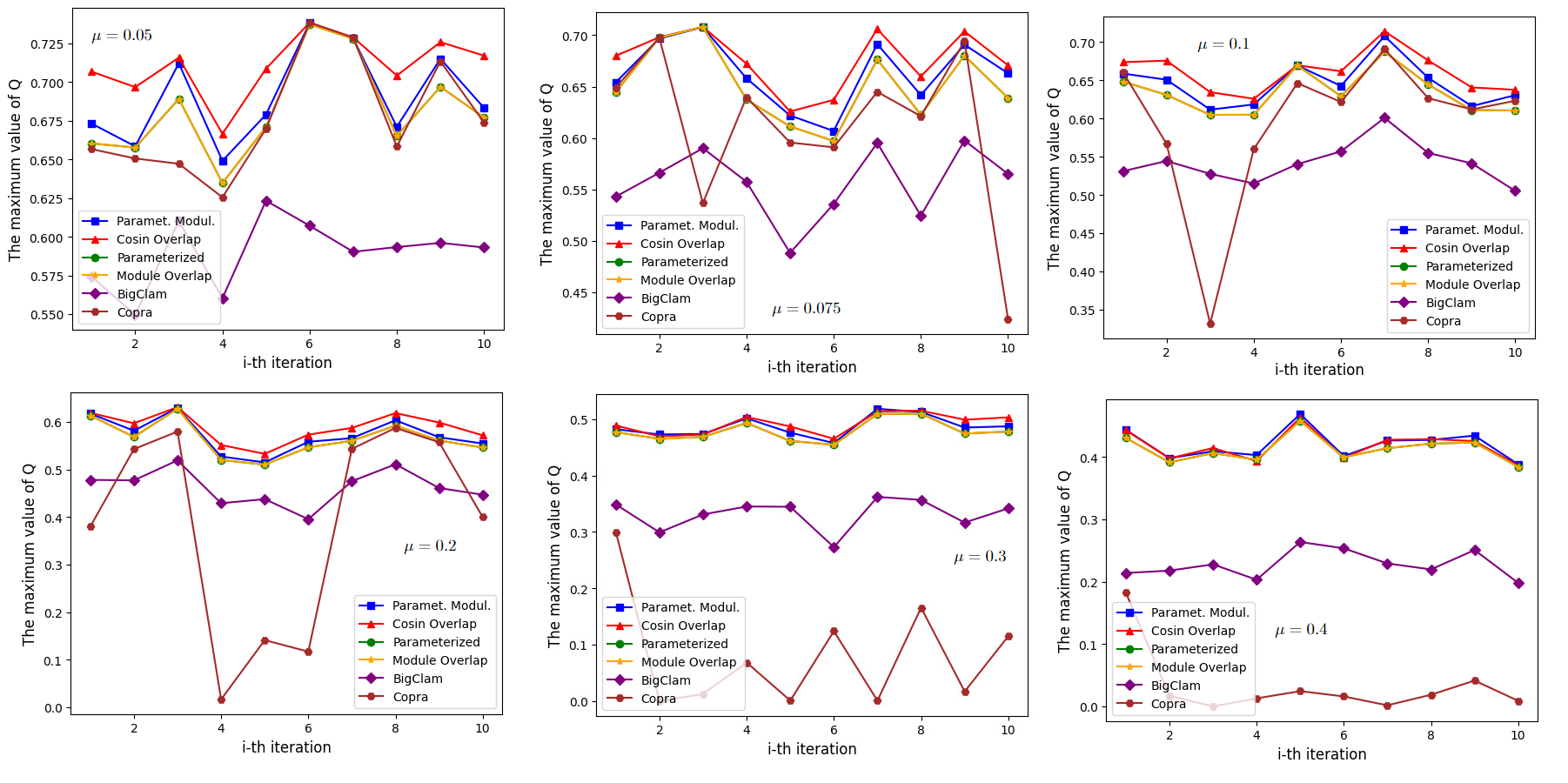}
\par\end{centering}
\caption{This chart illustrates the maximum modularity obtained in \textbf{Experiment 1 for undirected graphs}; we experimented on ten randomly generated graphs using the LFR benchmark with all the parameters taken with a uniform distribution in the following corresponding intervals: $N \in [400;500]$, $on \in [60;80]$, $om \in [2;5]$. }\label{Fig4a}
\end{figure}

\begin{figure}
\begin{centering}
\includegraphics[width=1\columnwidth]{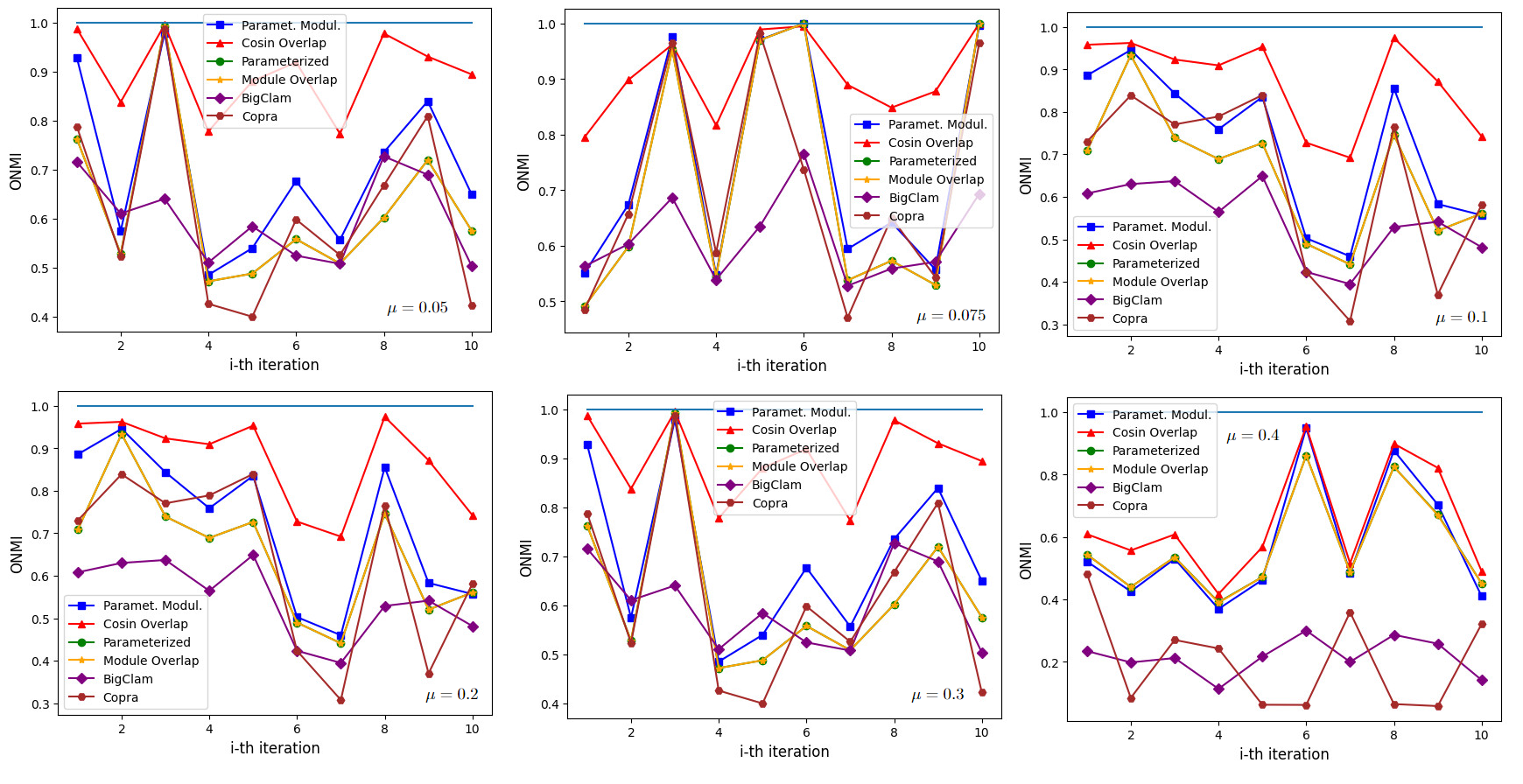}
\par\end{centering}
\caption{This chart illustrates the maximum ONMI obtained in \textbf{Experiment 1 for undirected graphs}; we experimented on ten randomly generated graphs using the LFR benchmark with all the parameters taken with a uniform distribution in the following corresponding intervals: $N \in [400;500]$, $on \in [60;80]$, $om \in [2;5]$. }\label{Fig4b}
\end{figure}

We have illustrated the results obtained from this experiment as shown in Figures \ref{Fig4a}, \ref{Fig4b}. In the scenario where a graph consists of approximately 400 to 500 vertices, with an overlap of 60 to 80 vertices, our two algorithms still have been shown to yield the maximum modularity value and ONMI value in most cases. Among these algorithms, the Cosine Overlap Algorithm has the best results.

\subsubsection*{Experiment 2 for undirected graphs:} We will experiment on ten randomly generated graphs using the LFR benchmark with all the parameters taken with a uniform distribution in the following corresponding intervals: $N \in [800;1000]$, $k \in [6;10]$, $maxk\in [10;20]$, $on \in [80;100]$, $om \in [2;5]$, $minc\in [70;100]$, and $maxc \in [100;150]$.

\begin{figure}
\begin{centering}
\includegraphics[width=1\columnwidth]{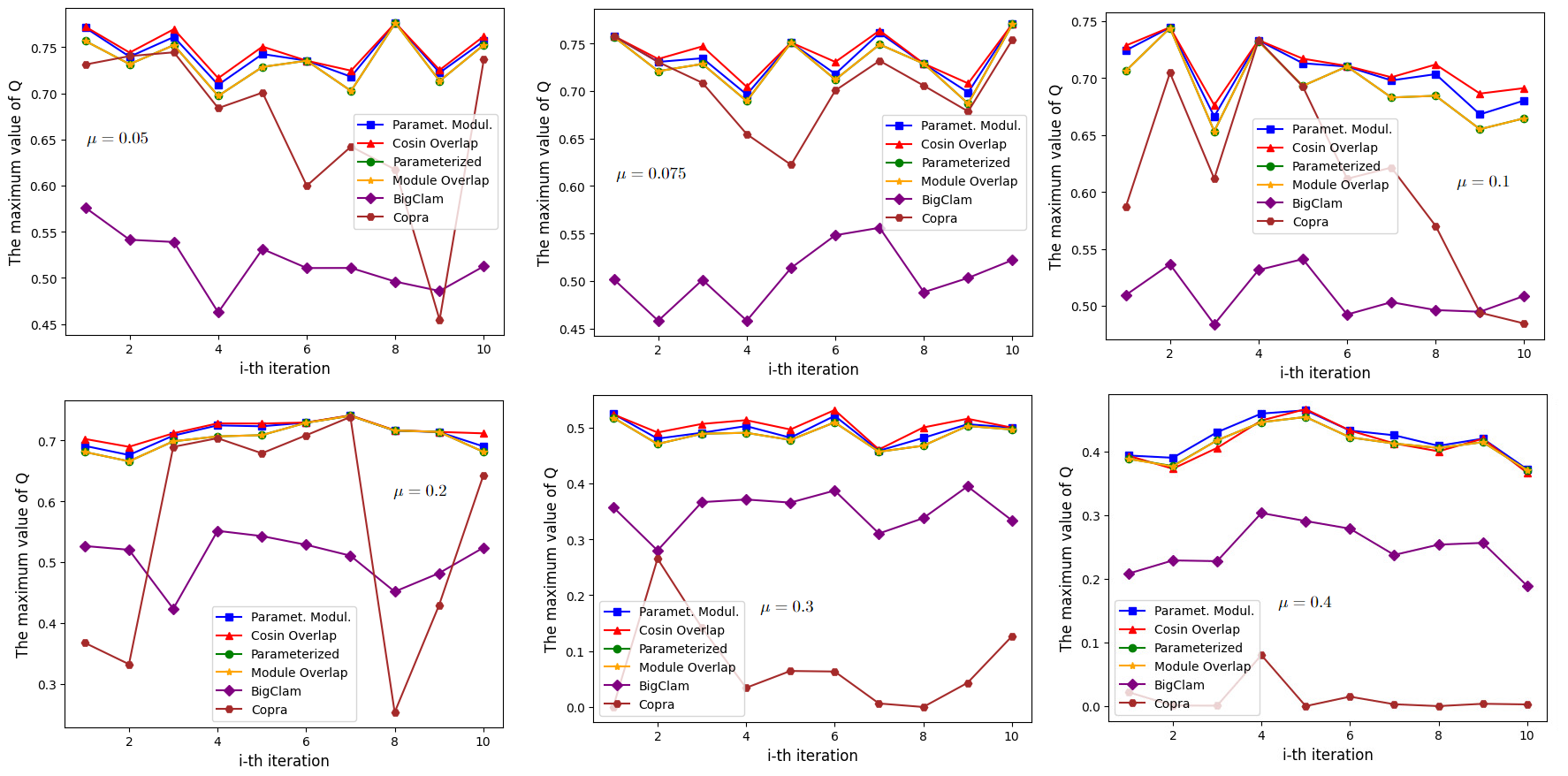}
\par\end{centering}
\caption{This chart illustratesthe maximum modularity obtained in \textbf{Experiment 2 for undirected graphs}; we experimented on ten randomly generated graphs using the LFR benchmark with all the parameters taken with a uniform distribution in the following corresponding intervals: $N \in [800;1000]$, $on \in [80;100]$, $om \in [2;5]$. }\label{Fig5a}
\end{figure}

\begin{figure}
\begin{centering}
\includegraphics[width=1\columnwidth]{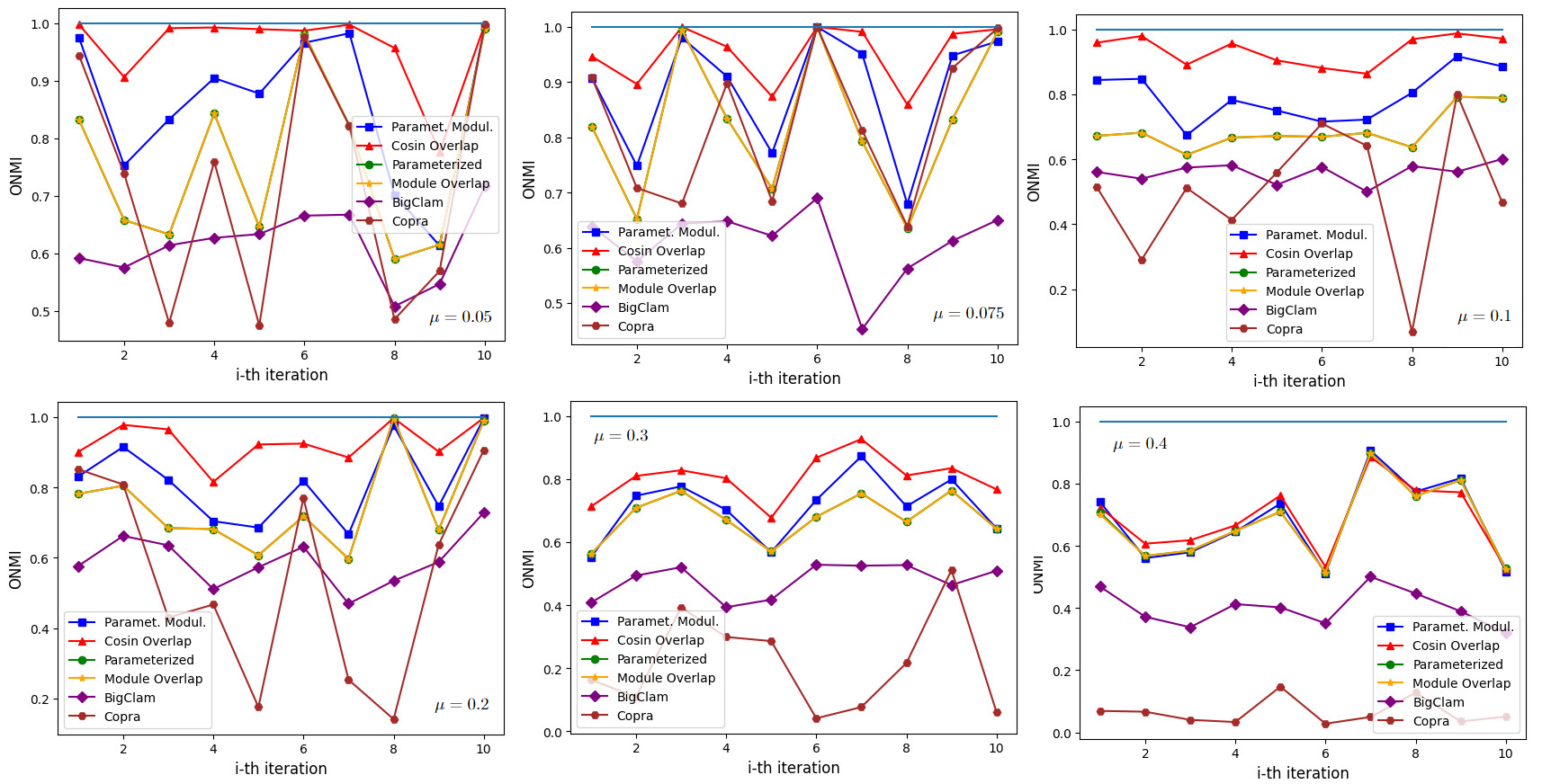}
\par\end{centering}
\caption{This chart illustratesthe maximum ONMI obtained in \textbf{Experiment 2 for undirected graphs}; we experimented on ten randomly generated graphsusing the LFR benchmark with all the parameters taken with a uniform distribution in the following corresponding intervals: $N \in [800;1000]$, $on \in [80;100]$, $om \in [2;5]$. }\label{Fig5b}
\end{figure}

Despite increasing the number of peaks in the graphs to approximately 800 to 1000, we could replicate the same results from the previous two experiments. We present these results in  Figures \ref{Fig5a}, \ref{Fig5b}.

\subsubsection*{Experiment 3 for undirected graphs:} We will experiment on ten randomly generated graphs using the LFR benchmark with all the parameters taken with a uniform distribution in the following corresponding intervals: $N \in [2000;3000]$, $k \in [10;20]$, $maxk\in [20;30]$, $on \in [200;300]$, $om \in [2;5]$, $minc\in [200;300]$, and $maxc \in [300;500]$.

\begin{figure}
\begin{centering}
\includegraphics[width=1\columnwidth]{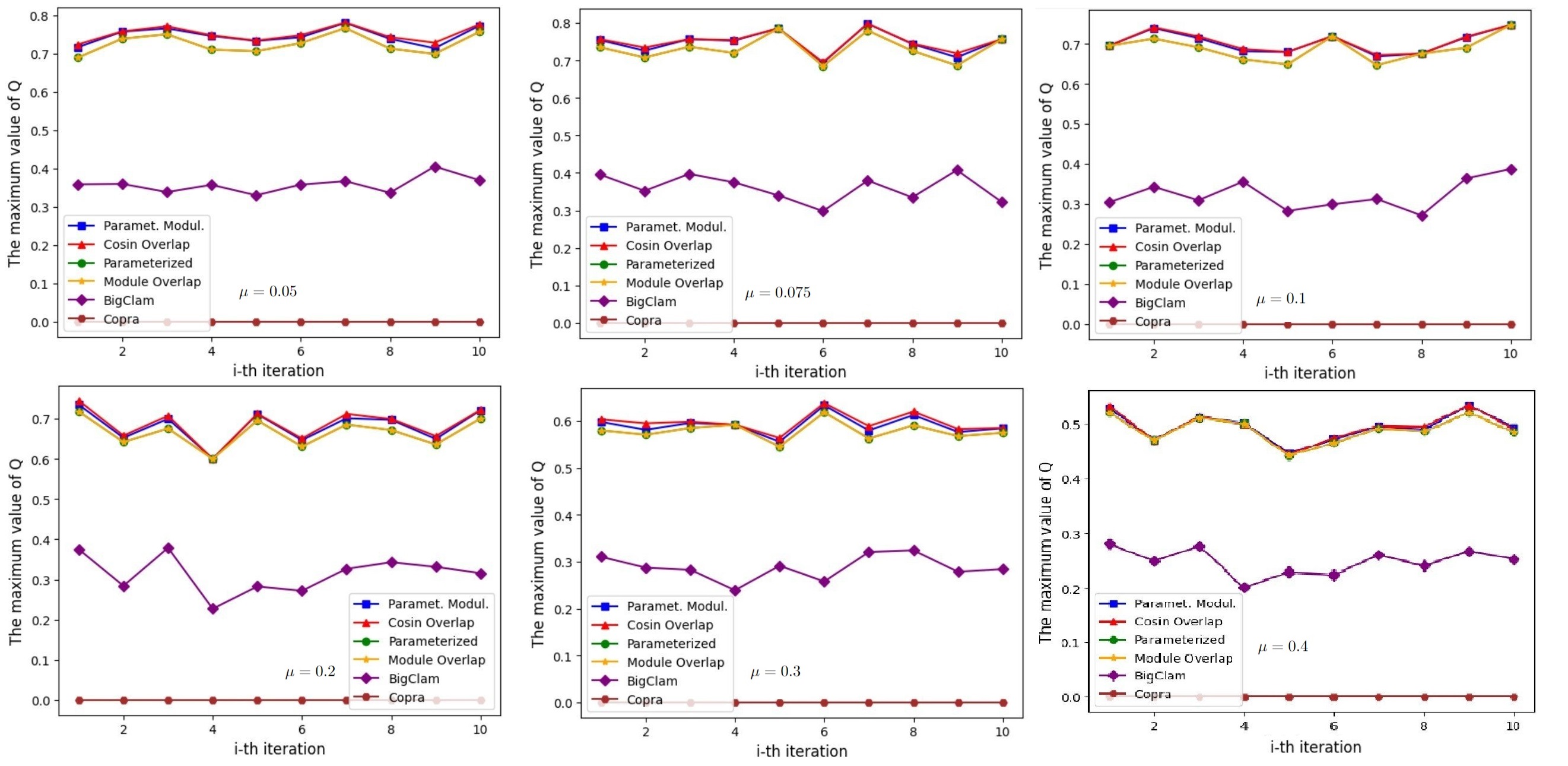}
\par\end{centering}
\caption{This chart illustratesthe maximum modularity obtained in \textbf{Experiment 3 for undirected graphs}; we experimented on ten randomly generated graphs using the LFR benchmark with all the parameters taken with a uniform distribution in the following corresponding intervals: $N \in [2000;3000]$,  $on \in [200;300]$, $om \in [2;5]$.}\label{Fig6a}
\end{figure}

\begin{figure}
\begin{centering}
\includegraphics[width=1\columnwidth]{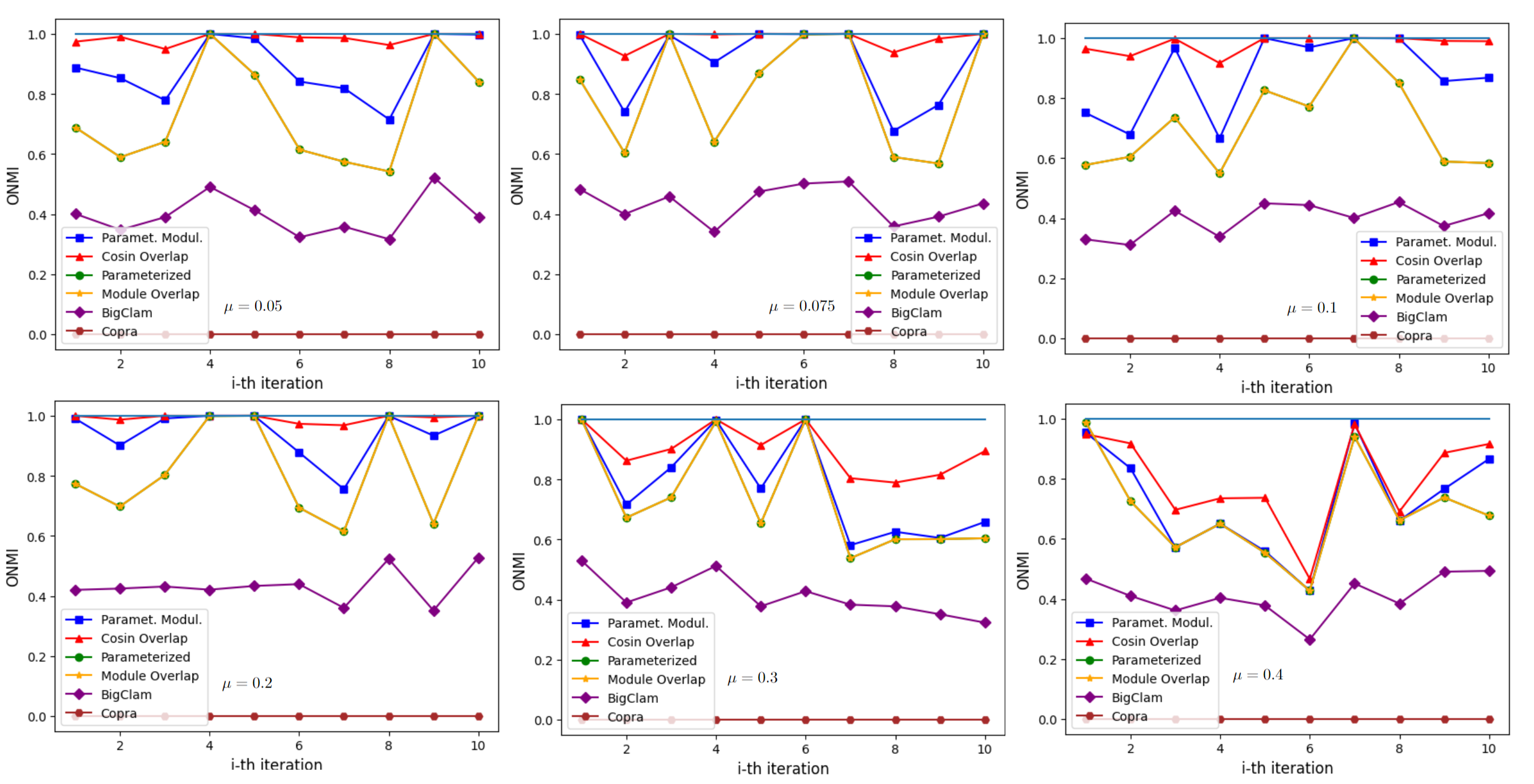}
\par\end{centering}
\caption{This chart illustratesthe maximum ONMI obtained in \textbf{Experiment 3 for undirected graphs}; we experimented on ten randomly generated graphs using the LFR benchmark with all the parameters taken with a uniform distribution in the following corresponding intervals: $N \in [2000;3000]$,  $on \in [200;300]$, $om \in [2;5]$.}\label{Fig6b}
\end{figure}

Despite increasing the number of peaks in the graphs to approximately 800 to 1000, we could replicate the same results from the previous two experiments. We present these results in  Figures \ref{Fig6a}, \ref{Fig6b}.

\subsubsection*{Experiment 4 for undirected graphs:} We will experiment on ten randomly generated graphs using the LFR benchmark with all the parameters taken with a uniform distribution in the following corresponding intervals: $N \in [5000;10000]$, $k \in [10;20]$, $maxk\in [20;30]$, $on \in [200;300]$, $om \in [2;5]$, $minc\in [200;300]$, and $maxc \in [300;500]$.

\begin{figure}
\begin{centering}
\includegraphics[width=1\columnwidth]{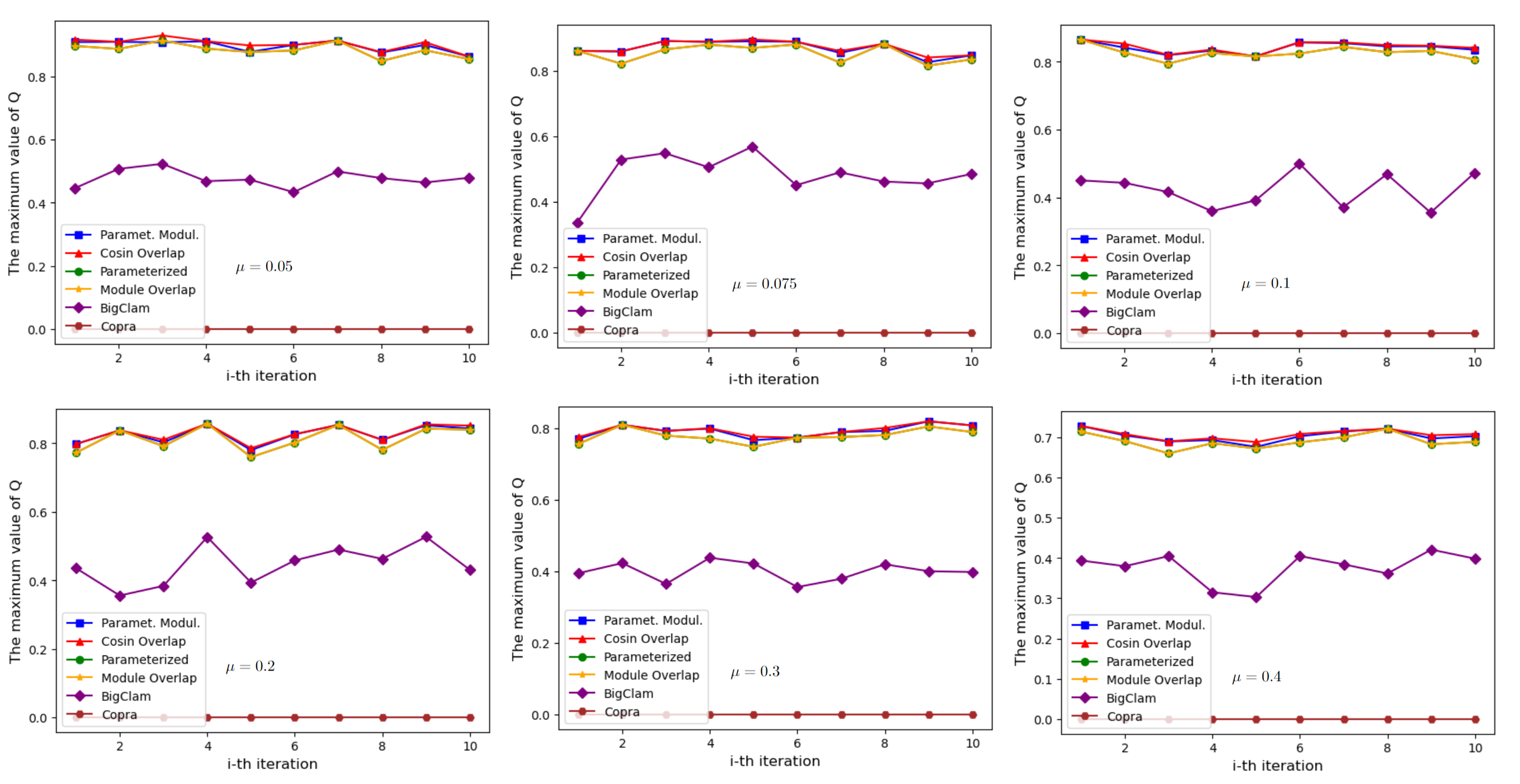}
\par\end{centering}
\caption{This chart illustratesthe maximum modularity obtained in \textbf{Experiment 4 for undirected graphs}; we experimented on ten randomly generated graphs using the LFR benchmark with all the parameters taken with a uniform distribution in the following corresponding intervals: $N \in [5000;10000]$, $on \in [200;300]$, $om \in [2;5]$.}\label{Fig7a}
\end{figure}

\begin{figure}
\begin{centering}
\includegraphics[width=1\columnwidth]{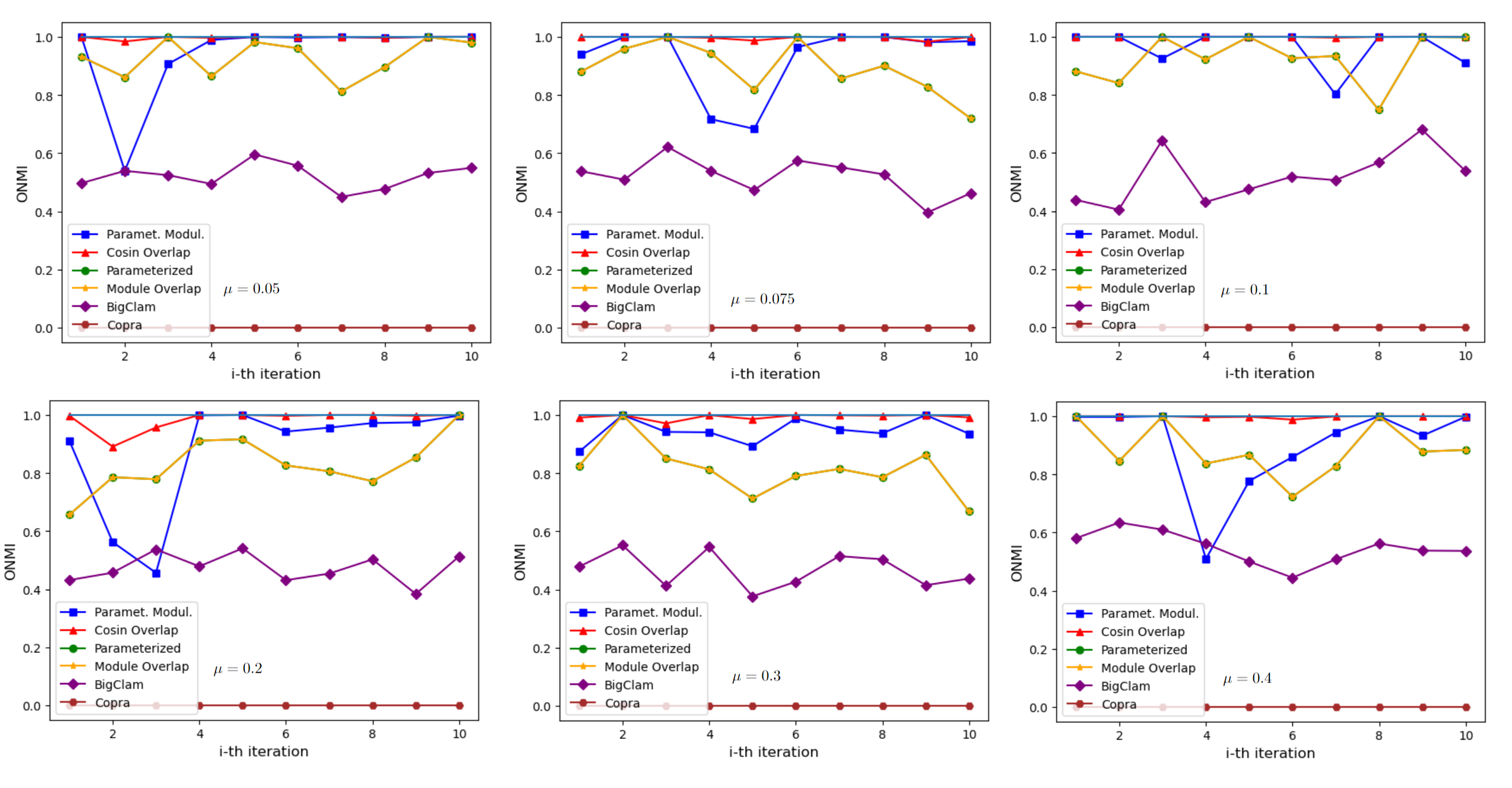}
\par\end{centering}
\caption{This chart illustratesthe maximum ONMI obtained in \textbf{Experiment 4 for undirected graphs}; we experimented on ten randomly generated graphs using the LFR benchmark with all the parameters taken with a uniform distribution in the following corresponding intervals: $N \in [5000;10000]$, $on \in [200;300]$, $om \in [2;5]$. }\label{Fig7b}
\end{figure}

Despite increasing the number of peaks in the graphs to approximately 5000 to 10000, we could replicate the same results from the previous two experiments. We present these results in  Figures \ref{Fig7a}, \ref{Fig7b}.

\subsubsection{Experiments on random graphs for directed graphs}\label{E_cohuong}
In the following experiments, we will utilize our three algorithms (Directed Parameterized d-Modularity Overlap Algorithm, Directed Parameterized sd-Modularity Overlap Algorithm, and Directed Cosine Overlap Algorithm) for directed graphs to cluster random generation graphs. Subsequently, we will compare the index obtained from our algorithms with the graph generation index. The parameters for each Algorithm will be as follows:
\begin{itemize}
    \item Directed Parameterized d-Modularity Overlap Algorithm: we will use the coefficient $\theta=1+0.1t$ with $t\in\{1,2,\dots,20\}$.
    \item Directed Parameterized sd-Modularity Overlap Algorithm: we will use the coefficient $\theta=1+0.1t$ with $t\in\{1,2,\dots,20\}$.
 \item Directed Cosine Overlap Algorithm: we will use the coefficient $\theta=0.2+0.035t$ with $t\in\{1,2,\dots,20\}$.
\end{itemize}
We will compare the clustering results obtained by each algorithm with the original clustering generated by the graph generation (original clustering) based on ONMI (the ONMI we use is the Formula \ref{ONMI}). ONMI measures the similarity between two sets of clusters, with a value close to 1 indicating a strong resemblance between the obtained and original clustering. When the obtained clustering matches the original clustering, the ONMI value will be 1. We will select the clustering result for each algorithm corresponding to the parameter value that obtained the highest ONMI value.

\subsubsection*{Experiment 1 for directed graphs:}  We will experiment on ten randomly generated graphs using the LFR benchmark with all the parameters taken with a uniform distribution in the following corresponding intervals: $N \in [400; 500]$, $k \in [6; 10]$, $maxk\in [10; 15]$, $on \in [20;30]$, $om \in [2;5]$, $minc\in [30; 50]$, and $maxc \in [50; 80]$.

\begin{figure}
\begin{centering}
\includegraphics[width=1\columnwidth]{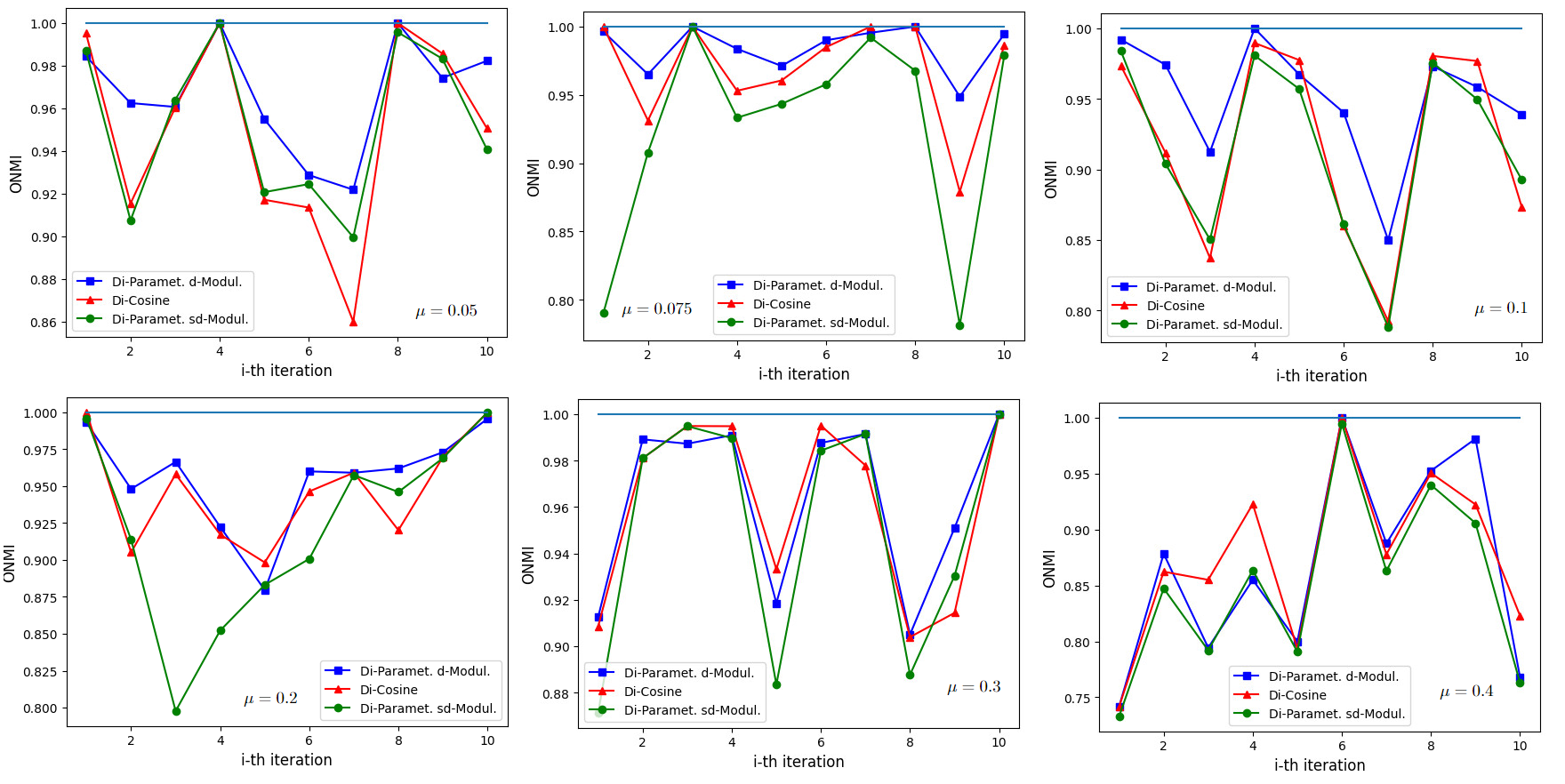}
\par\end{centering}
\caption{This chart illustrates the maximum ONMI obtained in \textbf{Experiment 1 for directed graphs}; we experimented on ten randomly generated graphs  using the LFR benchmark with all the parameters taken with a uniform distribution in the following corresponding intervals: $N \in [400; 500]$, $on \in [20;30]$, $om \in [2;5]$.}\label{Exp2d}
\end{figure}

 This experiment will investigate the number of overlapping vertices identified by the algorithms and compare it with those generated by the graph generation method. That will enable us to evaluate the efficiency of our Algorithm. We present the results of Experiment 1  in Figure \ref{Exp2d}.
   
\subsubsection*{Experiment 2 for directed graphs:}  We will experiment on ten randomly generated graphs using the LFR benchmark with all the parameters taken with a uniform distribution in the following corresponding intervals: $N \in [800; 1000]$, $k \in [8; 15]$, $maxk\in [15; 20]$, $on \in [40; 50]$, $om \in [2;5]$, $minc\in [80; 100]$, and $maxc \in [100; 150]$.

\begin{figure}
\begin{centering}
\includegraphics[width=1\columnwidth]{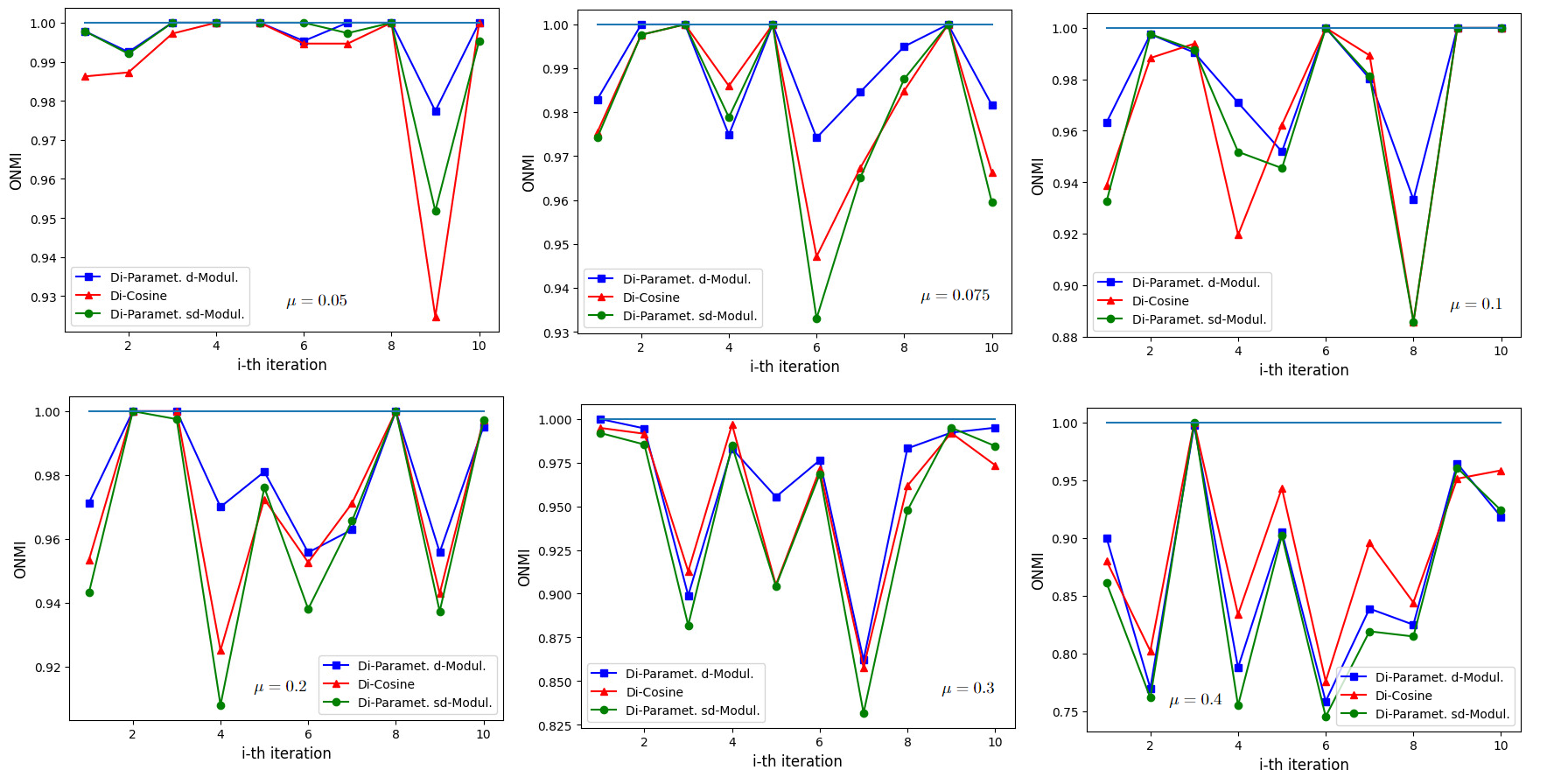}
\par\end{centering}
\caption{This chart illustrates the maximum ONMI obtained in \textbf{Experiment 2 for directed graphs}; we experimented on ten randomly generated graphs  using the LFR benchmark with all the parameters taken with a uniform distribution in the following corresponding intervals: $N \in [800; 1000]$, $on \in [40; 50]$, $om \in [2;5]$.}\label{Exp3d}
\end{figure}

 This experiment will investigate the number of overlapping vertices identified by the algorithms and compare it with those generated by the graph generation method. That will enable us to evaluate the efficiency of our Algorithm. We present the results of Experiment 2  in Figure \ref{Exp3d}.

\subsubsection*{Experiment 3 for directed graphs:}  
 We will experiment on ten randomly generated graphs using the LFR benchmark with all the parameters taken with a uniform distribution in the following corresponding intervals: $N \in [2000; 3000]$, $k \in [10; 20]$, $maxk\in [20; 30]$, $on \in [100; 150]$, $om \in [2;5]$, $minc\in [100; 200]$, and $maxc \in [300; 500]$.

\begin{figure}
\begin{centering}
\includegraphics[width=1\columnwidth]{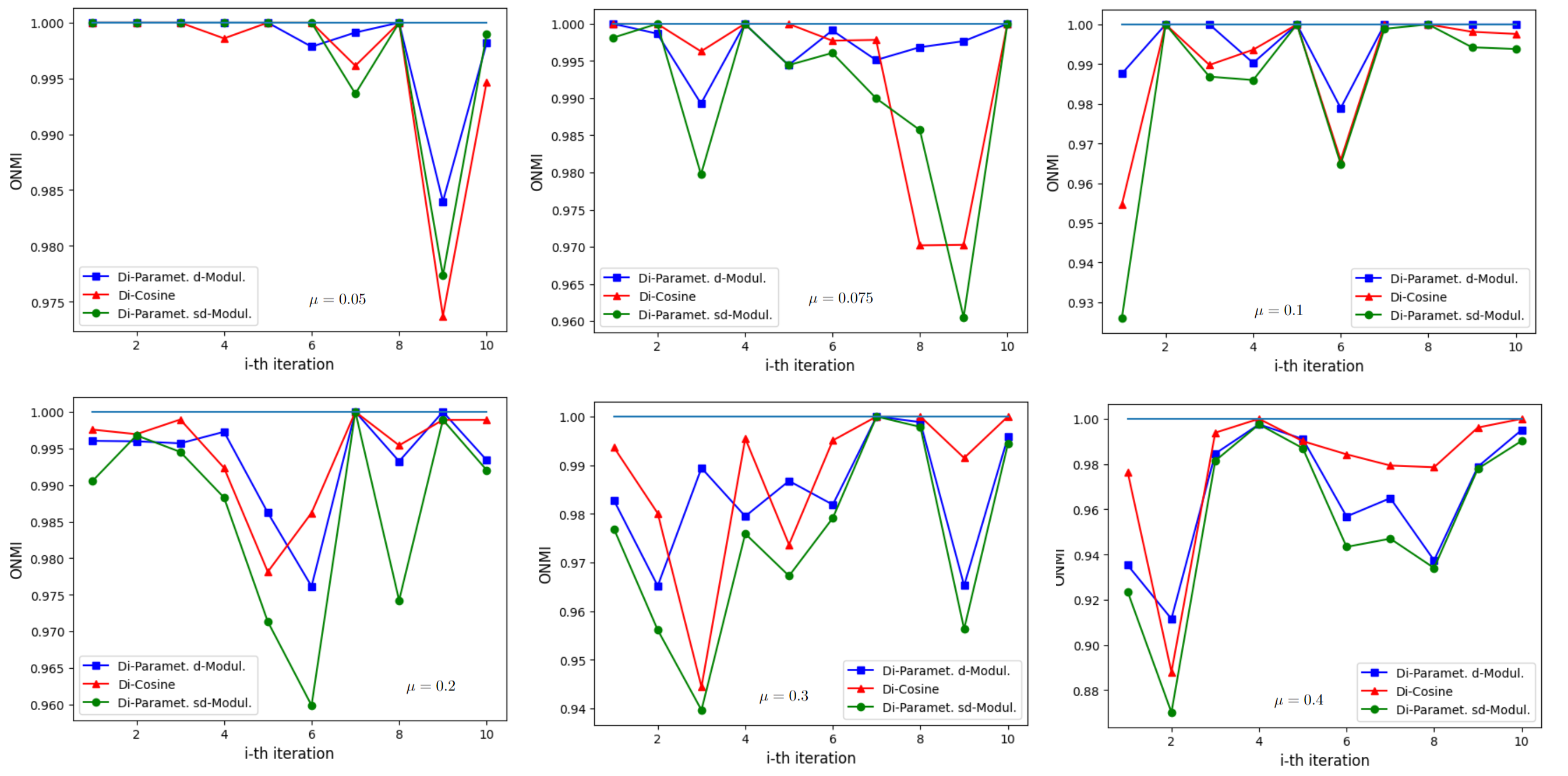}
\par\end{centering}
\caption{This chart illustrates the maximum ONMI obtained in \textbf{Experiment 4 for directed graphs}; we experimented on ten randomly generated graphs using the LFR benchmark with all the parameters taken with a uniform distribution in the following corresponding intervals: $N \in [2000; 3000]$, $on \in [100; 150]$, $om \in [2;5]$.}\label{Exp4d}
\end{figure}

 This experiment will investigate the number of overlapping vertices identified by the algorithms and compare it with those generated by the graph generation method. That will enable us to evaluate the efficiency of our Algorithm. We present the results of Experiment 3  in Figure \ref{Exp4d}.

\subsubsection*{Experiment 4 for directed graphs:}  
 We will experiment on ten randomly generated graphs using the LFR benchmark with all the parameters taken with a uniform distribution in the following corresponding intervals: $N \in [5000; 10000]$, $k \in [10; 30]$, $maxk\in [30;50]$, $on \in [100; 150]$, $om \in [2;5]$, $minc\in [100; 200]$, and $maxc \in [300; 500]$.

\begin{figure}
\begin{centering}
\includegraphics[width=1\columnwidth]{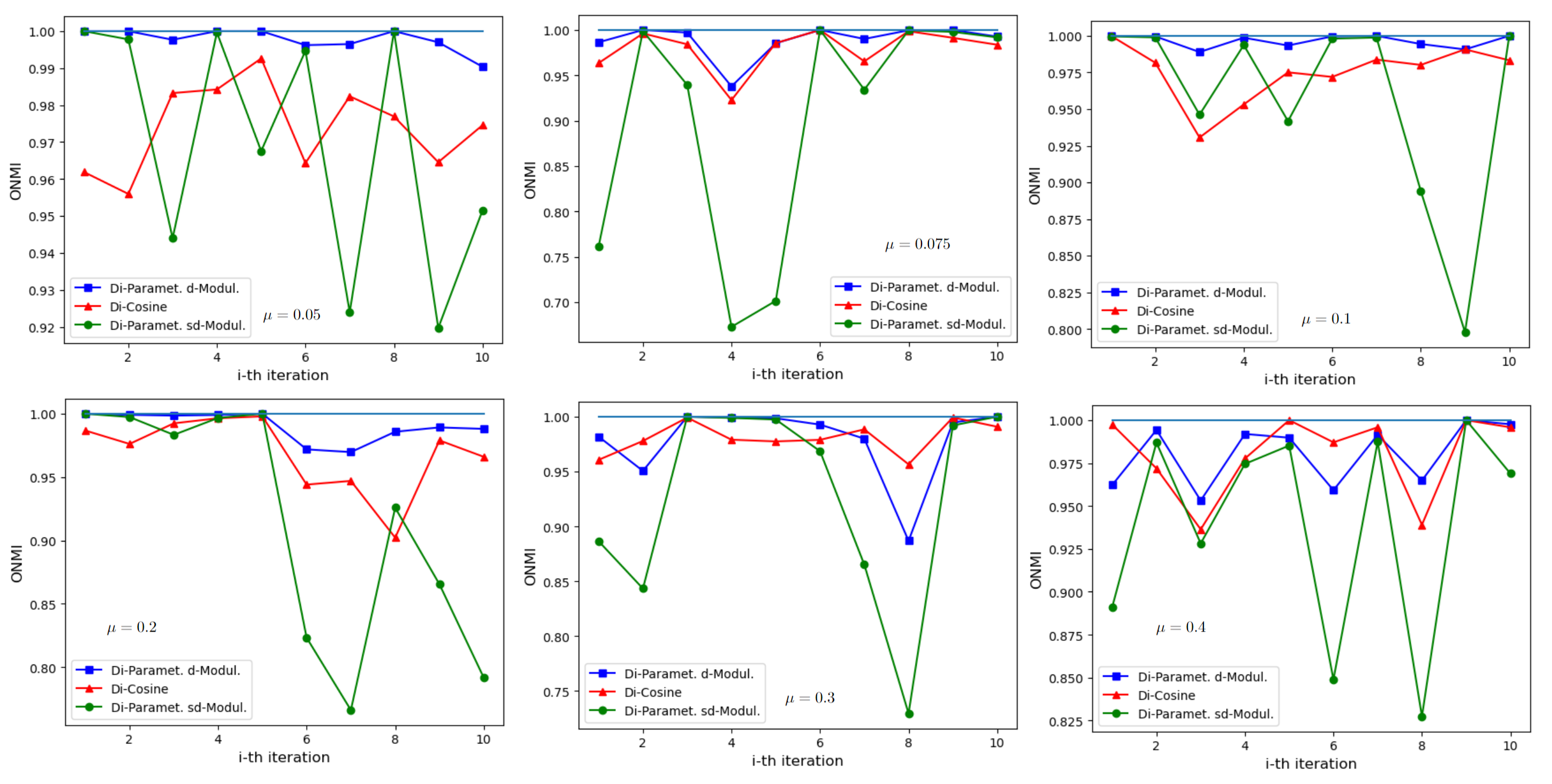}
\par\end{centering}
\caption{This chart illustrates the maximum ONMI obtained in \textbf{Experiment 4 for directed graphs}; we experimented on ten randomly generated graphs using the LFR benchmark with all the parameters taken with a uniform distribution in the following corresponding intervals: $N \in [5000; 10000]$, $on \in [100; 150]$, $om \in [2;5]$.}\label{Exp5d}
\end{figure}

 This experiment will investigate the number of overlapping vertices identified by the algorithms and compare it with those generated by the graph generation method. That will enable us to evaluate the efficiency of our Algorithm. We present the results of Experiment 4  in Figure \ref{Exp5d}.

\subsection{Real data and experiments on real data}
\subsubsection{Real data}
In this paper, we will perform experiments on the following famous real data. \\
\textbf{Zachary's karate club:} Wayne W. Zachary studied a social network of a karate club over three years from 1970 to 1972, as a paper in \cite{kara}. The network represents 34 members, recording the connections between pairs of members who had interactions beyond the club's premises. After being utilized by Michelle Girvan and Mark Newman in 2002 \cite{Foot}, the network became a widely-used example of community structure in networks.\\
\textbf{Dolphin’s associations:}
The dataset used in this study was obtained from \cite{Do}. It describes the connections between 62 dolphins living in Doubtful Sound, New Zealand, where the links between pairs of dolphins represent statistically significant frequent associations. This network can be naturally divided into two distinct groups. \\
\textbf{Metabolic network:}
According to \cite{meta}, a metabolic network represents the comprehensive collection of metabolic and physical processes that dictate a cell's physiological and biochemical characteristics. Therefore, these networks consist of metabolic reactions, pathways, and the regulatory interactions that direct these reactions.

In addition, we will also perform experiments on the following real data: Jazz network in \cite{Ncite7}. Email network \cite{email}. College football in \cite{Foot}. Hamster households, hamster friendships, DNC co-recipient, and Asoiaf in \cite{hams}.

\subsubsection{Experiments on real data}

We will conduct experiments on all six algorithms for each real network. We will perform 20 experiments with 20 different parameters for each Algorithm and select the clustering result with the highest Modularity among those experiments. Specifically, the parameters for each Algorithm will be set as follows:
\begin{itemize}
    \item 
Parameterized Modularity Overlap Algorithm: we will use the coefficient $\theta=1+0.1t$ with $t\in\{1,2,\dots,20\}$.
 \item Cosine Overlap Algorithm: we will use the coefficient $\theta=0.2+0.035t$ with $t\in\{1,2,\dots,20\}$.
 \item Parameterized Overlap Algorithm: we will use the coefficient $\theta=0.2+0.015t$ with $t\in\{1,2,\dots,20\}$.
 \item Module Overlap Algorithm: we will set the coefficient $B^{U}$ to 0.5 and use the coefficient $B^{L}=0.2+0.015t$ with $t\in\{1,2,\dots,20\}$.
  \item Copra Algorithm: We will execute the algorithm with each graph nine times, varying the parameter $V$ from 2 to 10 and keeping the parameter $T$ fixed at 15. After running these nine experiments, we will select the best result obtained among these runs. We will do this 20 times and choose the best result.
  \item Bigclam Algorithm: In each experiment, we will conduct form 5 to 20 runs of the algorithm, each with a different value for the parameter $K$ (number of communities). After running these 20 experiments, we will select the best result obtained from these runs. The values of the $K$ parameter will be chosen based on the size of the networks. For instance, in the case of the Karate network and the Dolphin network, we will consider values of $K$ ranging from 2 to 6.
\end{itemize}

\begin{table} 
\begin{centering}
\begin{tabular}{|p{2cm}|p{2cm}|p{2cm}|p{2cm}|p{2cm}|p{2cm}|p{2cm}|}
\hline Graph, $G=(|V|,|E|)$ &  Paramet. Modul. overlap& Cosine Overlap& Paramet. Overlap & Module Overlap  &  BigClam & Copra  \tabularnewline
\hline 
\hline 
Dolphin’s associations  \cite{Do}, $G = ( 62, 159 )$ & 0.5318953 &  0.5309415 &  0.5282947   &  0.5282947&0.4138026  & 0.4276747 \tabularnewline
\hline 
 Zachary’s karate club \cite{kara}, $G = ( 34, 78 )$ &  0.4300419  & 0.4197896  & 0.4267998  & 0.4267998  & 0.2741145 & 0.3920940 \tabularnewline
\hline 
 Metabolic network \cite{meta}, $G = ( 453 , 2025 )$ &  0.4542220  &  0.4385402  &  0.4501564 &  0.4447795  &0.2680016  &  0.0402168 \tabularnewline
\hline 
  College football \cite{Foot}, $G = ( 115, 613 )$&  0.6103202 & 0.6077106   & 0.6056203  & 0.6044072  & 0.5730757 & 0.5913225 \tabularnewline
\hline 
  Jazz network \cite{Jaz}, $G = ( 198, 2742 )$ &  0.4551599  & 0.4498614 &  0.4473198  & 0.4463121   &  0.3641616& 0.2528598 \tabularnewline
\hline 
 Email network \cite{email}, $G = ( 1133, 5451 )$ & 0.5900769   & 0.5905986 & 0.5900962       &    0.5786008 & 0.3961421 & 0.4997626\tabularnewline
\hline 
 Hamster households \cite{hams}, $ G = ( 921, 4032 )$ &  0.3804496 & 0.3658147   &  0.3782052  &  0.3774382  & 0.1210200 & 0.0293135  \tabularnewline
\hline 
Hamsters friendships \cite{hams}, $G = ( 1858, 12534 )$  & 0.4692017 & 0.4512992   & 0.4537756  &  0.4547541 &0.3424725  &  0.2063043 \tabularnewline
\hline 
 DNC co-recipient \cite{hams}, $G = ( 906 , 10429 )$  & 0.4434152   & 0.4420661   &  0.4419830 &  0.4419830   & 0.2838478  & 0.1822132 \tabularnewline
\hline 
Asoiaf \cite{hams}, $G = ( 796 , 2823 )$ &  0.6062354  &  0.6097880 & 0.609218 &  0.609218 & 0.4247218 &  0.5274229\tabularnewline
\hline 

\end{tabular}
\par\end{centering}
\caption{In this table, we present the values of modularity (using the formula \ref{Q_over}) corresponding to the clustering results of the algorithms.}\label{phan-data}

\end{table}

\subsection{Conclusion of the experiments}
The above results show that our two algorithms are efficient in almost all experiments.
\begin{itemize}
    \item  The Parameterized Modularity Overlap Algorithm consistently outperforms the Parameterized Overlap Algorithm, the Module Overlap Algorithm, Bigclam Algorithm, and Copra Algorithm in most experiments. Additionally, this algorithm offers the advantage of having low computational complexity.
\item  We built The Cosine Overlap Algorithm by observing the network's relationship between random walks and community structure. Although the computational complexity will be greater than the Parameterized Modularity Overlap Algorithm, all experiments on the Cosine algorithm randomization graph are for the best results. Furthermore, the Algorithm makes a lot of sense in theory and is an interesting algorithm that deserves attention.
\item Our algorithms for undirected graphs exhibit better performance compared to the other four algorithms, particularly when each overlapping vertex belongs to more than two communities. Additionally, in the case of graphs with a clear community structure, Cosine Overlap Algorithm consistently outperforms all other algorithms, producing superior results.
\item Our directed graph algorithms demonstrate remarkable efficiency compared to the indices generated based on the graph generation method. Especially the Directed Cosine Overlap Algorithm, in most cases, gets the best results.

\end{itemize}

\section{Conclusion and further work}

In this paper, we have proposed two algorithms for overlapping community detection for undirected and directed graphs; our algorithms go through 2 steps. In step 1, we separate community detection using the algorithms we know, such as the Hitting times Walktrap algorithm, NL-PCA 
 algorithm\cite{cohuong}, Walktrap algorithm \cite{Walktrap} or Louvain algorithm \cite{Louvain}. In step 2, we look for overlapping communities. Specifically, we have proposed the following two algorithms.
\begin{itemize}
\item  The Parameterized Modularity Overlap Algorithm uses the idea that vertex  $u$ belongs to the community $C_j$ if the sum of the probabilities from vertex  $u$ to the community $C_j$ and the probabilities from community $C_j$ to vertex  $u$ is large enough.
\item In the Cosine Overlap Algorithm, we first coordinate the vertices of the graph, then find the centers of the clusters and use the idea that the vertex  $u$ belongs to the cluster $C_j$ if the angle between the vector corresponds to the vertex  $u$ and the center of the cluster $C_j$ is small.
\end{itemize}
We also performed experiments on random generative graphs and real data to compare with Module Overlap algorithm, Parameterized Overlap algorithm, Bigclam algorithm, and Copra algorithm; the result is that our algorithms give better results when using the modularity and ONMI measures.

In the future, we will continue to study the problem of finding overlapping communities based on two vertices belonging to a community through other criteria, such as using cut, distance, and cosine.

\section*{Acknowledgments}
This research was supported by the Institute of Mathematics, Vietnam Academy of Science and Technology, Project code: NVCC01.02/23-24.

\end{document}